\documentclass[prb,reprint,groupedaddress,showpacs,floatfix,superscriptaddress]{revtex4-2}
\usepackage[pdftex]{graphicx}
\usepackage{dcolumn}
\usepackage{bm}
\usepackage{amssymb}
\usepackage{amsmath}
\usepackage{algorithm}
\usepackage{algpseudocode}
\usepackage{array}
\usepackage{dcolumn}
\usepackage{footmisc}
\usepackage[utf8]{inputenc}
\usepackage{tikz}
\usetikzlibrary{shapes,arrows}
\usepackage{color}
\usepackage[colorlinks=true,
            linkcolor=blue,
            urlcolor=blue,
            citecolor=blue]{hyperref}
\usepackage{mathtools}
\usepackage{soul}
\usepackage{csquotes}
\usepackage{graphicx}

\graphicspath{{./mainFigs/}}

\begin{document}
\title{Insights from exact exchange-correlation kernels}
\author{N. D. Woods}
\email{nw361@cam.ac.uk}
\affiliation{Theory of Condensed Matter, Cavendish Laboratory, University of Cambridge, Cambridge, CB3 0HE, United Kingdom}
\author{M. T. Entwistle}
\thanks{Present address: FU Berlin, Department of Mathematics and Computer Science, Arnimallee 6, 14195 Berlin, Germany}
\affiliation{Department of Physics, University of York, and European Theoretical Spectroscopy Facility, Heslington, York YO10 5DD, United Kingdom}
\author{R. W. Godby}
\affiliation{Department of Physics, University of York, and European Theoretical Spectroscopy Facility, Heslington, York YO10 5DD, United Kingdom}

\date{\today}

\begin{abstract}
The exact exchange-correlation (xc) kernel $f_\text{xc}(x,x',\omega)$ of linear response time-dependent density functional theory is computed over a wide range of frequencies, for three canonical one-dimensional finite systems. Methods used to ensure the numerical robustness of $f_\text{xc}$ are set out. The frequency dependence of $f_\text{xc}$ is found to be due largely to its analytic structure, i.e. its singularities at certain frequencies, which are \textit{required} in order to capture particular transitions, including those of double excitation character. However, within the frequency range of the first few interacting excitations, $f_\text{xc}$ is approximately $\omega$-independent, meaning the \textit{exact adiabatic approximation} $f_\text{xc}(\omega = 0)$ remedies the failings of the local density approximation and random phase approximation for these lowest transitions. The key differences between the exact $f_\text{xc}$ and its common approximations are analyzed, and cannot be eliminated by exploiting the limited gauge freedom in $f_\text{xc}$. The optical spectrum benefits from using as accurate as possible an $f_\text{xc}$ and ground-state xc potential, while maintaining exact compatibility between the two is of less importance.
\end{abstract}

\maketitle

\section{Introduction}

The density-density linear response function of a many-body quantum system can be used to extract a great deal of excited-state information about the system, for example, its optical transition probabilities and transition energies when subject to incident light \citep{Martin2017}. Linear response time-dependent density functional theory (DFT) constitutes an exact methodology, in principle, for recovering an interacting response function from the response function of a corresponding Kohn-Sham system \citep{Ullrich2012,Marques2006,Ullrich2014,Burke2005,Casida2012,Marques2004,Gross1990}. 
The interacting $\chi(x,x',|t-t'|)$ density-density response function describes the first-order change in the density due to a perturbation in the external potential \citep{Ullrich2012,Marques2006}:
\begin{align}
\chi(x,x',|t-t'|) = \left. \frac{\delta n(x,t)}{\delta v_\text{ext}(x',t')} \right|_{n_0}.
\end{align}
The Kohn-Sham response function $\chi_0 = \delta n / \delta v_\text{KS}$ is specified with the ground-state exchange-correlation (xc) potential $v_\text{xc}(x)$, and then a map from the Kohn-Sham response function to the interacting response function is established using the \textit{xc kernel}, 
\begin{align}
f_{\text{xc}}(x,x',|t-t'|) = \left. \frac{\delta v_{\text{xc}}(x,t)}{\delta n(x',t')} \right|_{n_0}, \label{eq:xcKernelDef2}
\end{align}
i.e. the first-order change in the xc potential due to a perturbation in the density. The uniqueness of this map is guaranteed by the Runge-Gross theorem of time-dependent DFT \citep{Runge1984,VanLeeuwen1999}, and the definition of $f_\text{xc}$ in Eq$.$ (\ref{eq:xcKernelDef2}) demonstrates that, like the ground-state xc potential, $f_\text{xc}$ is a functional of the ground-state density $n_0$. The principal aim of this work is to elucidate the structure and features of the \textit{exact} numerical $f_\text{xc}$, that is, $f_\text{xc}(x,x',|t-t'|)$ including the full extent of its spatial and temporal character. 

The map from the Kohn-Sham response function to the interacting response function is identified with the requirement that density perturbations in the Kohn-Sham system match those in the interacting system.  This map is often referred to as the Dyson equation of linear response time-dependent DFT,
\begin{align}
\chi(x,x',\omega) =& \chi_0(x,x',\omega) + 
\iint \chi_0(x,x'',\omega) \{ f_\text{H}(x'',x''') \nonumber\\
&+ f_\text{xc}(x'',x''',\omega) \} \chi(x''',x',\omega) \ dx'' dx''', \nonumber
\end{align}
where $f_\text{H} = \delta v_\text{H} / \delta n$ is the Hartree kernel (the electron-electron interaction) and all objects are now expressed in the frequency domain $\omega$, the Fourier transform of the time domain $|t-t'|$. An approximate Kohn-Sham response function in conjunction with an approximate $f_\text{xc}$ provides an approximation to the exact interacting response function, from which a host of properties can be calculated, such as the optical absorption spectrum \citep{Onida2002}  and the ground-state correlation energy \citep{Gorling2019,Olsen2019}. The optical absorption spectrum \citep{Ullrich2012},
\begin{align}
\sigma(\omega) = -\frac{4 \pi \omega}{c} \iint \text{Im}(\chi(x,x',\omega)) x x' \ dxdx', \label{eq:opticalSpectrum}
\end{align}
is the main focus of this work, and provides the transition energies and transition rates of a sample subject to classical light within the dipole approximation \footnote{Hartree atomic units $m_e = \hbar = e = 4 \pi \varepsilon_0 = 1$ are used throughout.}. Linear response time-dependent DFT is now a prominent method used to excited- and ground-state aspects of finite and extended systems. 

The development and understanding of approximate xc kernels has been the subject of intense interest over the past decades, see, for example, \citep{Botti2007,Marques2004,Ullrich2012,Ullrich2016,Onida2002} and references therein. In most cases, these approximations can be sorted hierarchically depending on the level of theory involved in the approximation. The lowest orders of this hierarchy contain the random phase approximation (RPA) and the adiabatic local density approximation (LDA). The former ignores exchange and correlation \textit{at the level of the xc kernel} entirely by setting $f^\text{RPA}_\text{xc} = 0$ \citep{Gavrilenko1997}, and the latter includes exchange and correlation within the framework of an LDA, leading to an adiabatic, spatially local $f_\text{xc}^\text{ALDA} \propto \delta(x-x')\delta(t-t')$ \citep{Corradini1998,Moroni1995}. 

The xc kernel itself, however, is known to possess a range of pathological features that depart significantly from these approximations. In particular, certain circumstances demand a spatial ultra-non-locality in $f_\text{xc}$ \citep{Ullrich2012}. Furthermore, a non-adiabatic temporal structure is known to be essential to capture excitations of a multi-particle character \citep{Maitra2002,Maitra2004,Elliott2011,Cave2004}. These are two manifestations of the fact that the exact $f_\text{xc}$ contains all correlated many-body effects. More sophisticated approximations to $f_\text{xc}$ seek to include these effects in some form or another, such as those that utilize the $GW$ approximation and the Bethe-Salpeter equation \citep{Reining2002,Marini2003,Romaniello2009,Ullrich2016}, exact-exchange kernels \citep{Hellgren2009,Hellgren2008,YongHoon2002,Gorling1999,Gorling1998}, and long-range corrected kernels \citep{DelSole2003,Botti2004}. 

The use of model systems has been effective in developing understanding of $f_\text{xc}$. In particular, the frequency dependence of $f_\text{xc}$ has been the subject of model analytic studies \citep{Ruggenthaler2013,Maitra2004,Maitra2006,Gritsenko2009}, numerical studies using model Hamiltonians, e.g. the Hubbard model \citep{Aryasetiawan2002,Carrascal2018,Turkowski2014,Fuks2014}, and numerical studies of exact one-dimensional Hamiltonians in a truncated Hilbert space \citep{Thiele2009,Thiele2014, Entwistle2019}. This work continues along the lines of the last approach, and seeks to address the spatial \textit{and} frequency dependence of $f_\text{xc}$ for energies far beyond the first few excitations. The observed features of $f_\text{xc}$ are examined in relation to matters of practical interest, such as optical properties. 

\section{Methodology}

\subsection{Background}

The \verb|iDEA| code \citep{Hodgson2013} is used in order to obtain the interacting and Kohn-Sham response functions. This software implements quantum mechanics for finite systems in one dimension interacting with a softened Coulomb electron-electron interaction
\begin{align}
v_c(x,x') = \frac{1}{|x-x'| + \alpha}, \label{eq:softCoulomb}
\end{align}
where $\alpha$ is the extent of the softening; we use $\alpha = 1$ a.u. A delta function basis set, i.e. real-space grid, of dimension $N$ is used to discretize the spatial domain $[-a,a]$ of length $L=2a$ subject to Dirichlet boundary conditions.

Our three prototype systems each consist of two spinless electrons in the external potentials described below. (The use of spinless fermions delivers a richer ground state and excitation spectrum for a given number of electrons.)

For some input external potential $v_\text{ext}(x)$, the full set of eigenvectors $\{ | \Psi_i \rangle \}$ of the interacting Hamiltonian is found using exact diagonalization. The corresponding \textit{exact} Kohn-Sham potential $v_\text{KS}(x)$ is then reverse-engineered by applying preconditioned root-finding techniques to an appropriate fixed-point map \citep{Ruggenthaler2015}. The full set of Kohn-Sham eigenvectors $\{ | \phi_i \rangle \}$ is also obtained using exact diagonalization. The causal response functions are computed in the frequency domain directly using the Lehmann representation; the interacting response function, for example, is given by
\begin{align}
\chi(x,x',\omega) = \lim_{\eta \rightarrow 0} \sum_{n=1}^{\infty} \frac{\langle \Psi_0 | \hat{n}(x) | \Psi_n \rangle \langle \Psi_n | \hat{n}(x') | \Psi_0 \rangle}{\omega - \Omega_n + i \eta} \nonumber \\
- \frac{\langle \Psi_0 | \hat{n}(x') | \Psi_n \rangle \langle \Psi_n | \hat{n}(x) | \Psi_0 \rangle}{\omega + \Omega_n + i \eta}, \nonumber
\end{align}
where $\Omega_n = E_n - E_0$ is the $n^\text{th}$ excitation energy of the interacting Hamiltonian, and the response function is zero for times $t > t'$. Construction of the interacting response function in this fashion is an accurate but demanding procedure, whereas the methods outlined in \citep{Thiele2014} to construct the response functions are amenable to larger systems, but more prone to error; either method will suffice here. On a finite spatial grid, the response functions at a given $\omega$ become \textit{response matrices}, denoted $\chi(\omega)$ and $\chi_0(\omega)$. The Dyson equation gives an alternate definition of the xc kernel,
\begin{align}
f_\text{xc}(\omega) = \chi_0^{-1}(\omega) - \chi^{-1}(\omega) - f_\text{H}, \label{eq:xcKernel}
\end{align}
where $^{-1}$ is to be understood as the matrix inverse, and the Hartree kernel $f_\text{H}$ becomes softened due to Eq$.$ (\ref{eq:softCoulomb}). This expression is used as the definition of $f_\text{xc}$ in the present context, where the matrix inverses require the careful treatment described in the next section.

\subsection{Challenges in computing exact exchange-correlation kernels}
\label{sec:BackgroundNumericalChallenges}

Numerical difficulties arise when attempting to construct $f_\text{xc}$ as an object in itself using Eq$.$ (\ref{eq:xcKernel}), which is one of a few reasons that has prevented or hindered studies of the exact $f_\text{xc}$ along these lines \citep{Thiele2009,Thiele2014,Entwistle2019}. The xc kernel represents the solution to an \textit{inverse problem}, i.e. find the $\delta v_\text{xc}$ that produces a \textit{given} $\delta n$, and inverse problems are notoriously sensitive to small error \citep{Tarantola2005}, such as those introduced by finite-precision arithmetic. As discussed, $f_\text{xc}$ in Eq$.$ (\ref{eq:xcKernel}) requires the matrix inverse of the response matrices at a given $\omega$, and hence a naive inversion procedure introduces numerical error at a given $\omega$ in proportion to the condition of the response matrices, that is, the ratio of the maximum to minimum eigenvalue. 

Physical eigenvalues that are close to, or below, machine precision manifest in the response matrices from various sources \citep{VanLeeuwen2001}. One such source is related to the linear response $v$-representability problem. That is to say, there exist density perturbations that oscillate with some non-resonant frequency $\omega$ that cannot be produced by a perturbing potential at linear order \cite{Mearns1987,VanLeeuwen2001,Hellgren2009}. Therefore, at such a frequency, the response function has an eigenvector $|u(\omega)\rangle$ whose eigenvalue is zero, indicating that the density perturbation $\delta n = |u(\omega)\rangle$ does not correspond to some finite $\delta v$, as the response function is non-invertible \footnote{For example, the constant perturbation $\delta v = c(\omega)$ oscillating with frequency $\omega$ produces no response in the density $\delta n$ at \textit{all} orders, including at first order.}. This can happen in both the interacting and Kohn-Sham response functions at distinct frequencies. These eigenvalues are of importance for the work to follow, and are discussed in more depth in Section \ref{sec:InfPotWell}. 

Another source of low eigenvalues is due to extended regions of nearly vanishing ground-state density. Since the aim of this work is, in part, to study the optical response of \textit{confined} systems far beyond the first excitation, an appropriately large spatial domain $[-a,a]$ is required in order to accommodate the more extended excited states without introducing spurious features due to the boundary conditions. Within such a system, a perturbing potential localized toward the edge of the domain yields a negligible ground-state density response, the effect of which is to introduce near-zero eigenvalues into the response functions. Therefore, ill-conditioning is unavoidable if we are to study the response functions, and thus $f_\text{xc}$, at high frequencies. The extent of this ill-conditioning depends on the maximum frequency up to which one wishes to examine matters. 

Note that these near-machine-precision eigenvalues of the response functions are problematic only if this difference accounts for some particular physical phenomenon, such as charge transfer, whereby in order to capture excitations of charge-transfer character an $f_\text{xc}$ that is divergent in proportion to the increasing separation between the subsystems involved is required \footnote{We observed this situation in a double-well system that we explored as background to the present study.} \citep{Maitra2017}. In such cases the construction of numerical xc kernels will be challenging. However, the systems studied in this work do not suffer this issue: the ill-conditioning of the interacting and Kohn-Sham response functions can be assumed to cancel in the definition of $f_\text{xc}$, Eq$.$ (\ref{eq:xcKernel}), thus producing a regular $f_\text{xc}$. This procedure can be viewed  as a form of \textit{basis set truncation}, i.e. assign $\chi = \chi_0$ within some subset of the basis responsible for ill-conditioning and proceed to compute $f_\text{xc}$ under this assumption. We now describe two such approaches: a truncation in real space, and a truncation in eigenspace.

\subsubsection{Real-space truncation}
\label{sec:realSpaceTrunc}

In finite systems with a confining potential, the response functions tend toward zero outside of the confined region, and this so-called long-range behavior is known to be relatively unimportant in the present context -- this is not the case in periodic systems \citep{Ullrich2016,Botti2007,Ghosez1997,Byun2019}. Therefore, as we demonstrate in this work, forcing the interacting response function to equal the non-interacting response function within some yet undefined \textit{outer region} does not much alter the derived properties of the interacting response function, such as its optical spectrum. 

\begin{figure}[ht!]
\begin{center}
\includegraphics[width=3in]{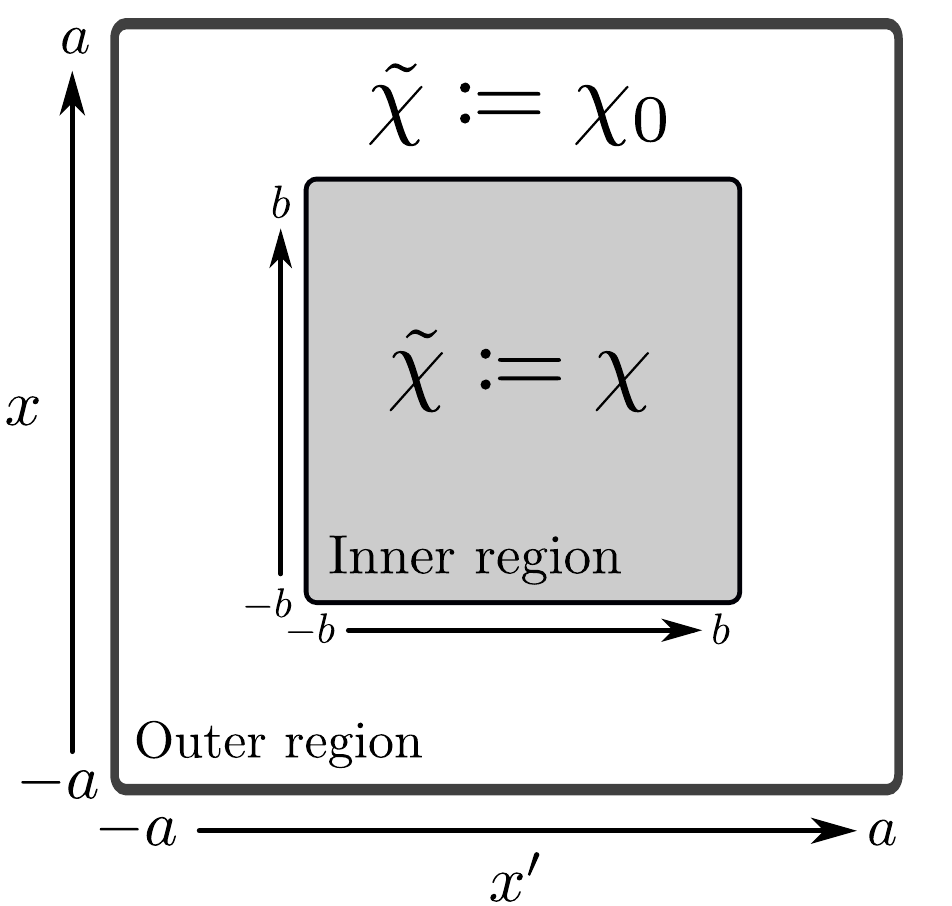}
\end{center}
\caption{A schematic depiction of the \textit{real-space truncation} strategy used to regularize computations of $f_\text{xc}$, whereby a truncated response function $\tilde{\chi}$ is defined as the interacting response function within some region parameterized by $b$, the \textit{inner region} (shaded gray), and is otherwise set equal to the Kohn-Sham response function.}
\label{fig:realSpaceTruncation}
\end{figure}

To this end, a partition of the spatial domain $[-a,a]$ is made such that an \textit{inner region} is defined where $x$ takes values $-b \leq x \leq b$; the numerical parameter $b$ defines the extent of the truncation. The \textit{outer region} constitutes the remaining space between the inner region and the edges of the domain, $-a$ and $a$. This partition of the space, as it applies to the response functions, can be seen in Fig$.$ \ref{fig:realSpaceTruncation}. The assumption is then made that
\begin{align}
\tilde{\chi}(x,x',\omega) &\coloneqq \chi(x,x',\omega) \text{ for } (x,x') \text{ in inner region} \nonumber \\
\tilde{\chi}(x,x',\omega) &\coloneqq \chi_0(x,x',\omega) \text{ for } (x,x') \text{ in outer region},  \nonumber
\end{align}
where $\tilde{\chi}$ is the \textit{truncated} response function.
The xc kernel is now defined as the object that returns the truncated response function $\tilde{\chi}$, rather than the interacting response function $\chi$, upon solution of the Dyson equation. This leads to the following piecewise form for $f_\text{xc}$,
\[
f_\text{xc} =
\begin{cases}
    \chi_0^{-1} - \chi^{-1} - f_{\text{H}} & \text{ for } (x,x') \text{ in inner region}  \\
    -f_{\text{H}} & \text{ for } (x,x') \text{ in outer region}.
\end{cases}
\]
The regularizing effect of the method presented above can be understood by examining the role of the truncation parameter $b$. In particular, two extremes are considered, first, setting $b=a$ means the inversion of the response matrix at a given $\omega$ is performed over the whole domain, and hence is dominated by error due to excessive regions of nearly vanishing ground-state density; this error is hereafter referred to as the \textit{numerical error}. Setting $b=0$ turns the truncated response function into the Kohn-Sham response function over the entire domain -- an evidently unsatisfactory state of affairs -- and error of this kind is referred to as \textit{method error}. Whilst this need not be the case in principle, it is the case for the systems studied here that it is possible to choose $b$ such that an acceptable balance is struck between method error and numerical error. In other words, the truncated response function is able to retain all the physical properties of the interacting response function \textit{and} ensure the computation of the resulting piecewise $f_\text{xc}$ is well-conditioned. A discussion on the notion of error in the present context, including an elaboration of the \textit{method error} and \textit{numerical error}, is given in the supplemental material \footnote{URL to be inserted}.

One might take issue that the above piecewise expression for $f_\text{xc}$ is spuriously discontinuous at the boundary of the inner and outer region, where the extent of this discontinuity depends on the long-range behavior of $f_\text{xc}$. However, we shall be concerned with the behavior of $f_\text{xc}$ inside the inner region, i.e. the region where departure of the Hxc kernel $f_\text{Hxc} = f_\text{xc} + f_\text{H}$ from zero produces meaningful features in the output of the Dyson equation. 

\subsubsection{Eigenspace truncation}

A second, related, method used in this work in order to regularize the computation of $f_\text{xc}$ is to truncate the interacting response matrix in the eigenspace of the Kohn-Sham response matrix. This method is much more accurate than the real-space truncation, but is limited to Hermitian response matrices, i.e. response matrices constructed without an artificial broadening $\eta$. 

\begin{figure}[ht!]
\begin{center}
\includegraphics[width=3in]{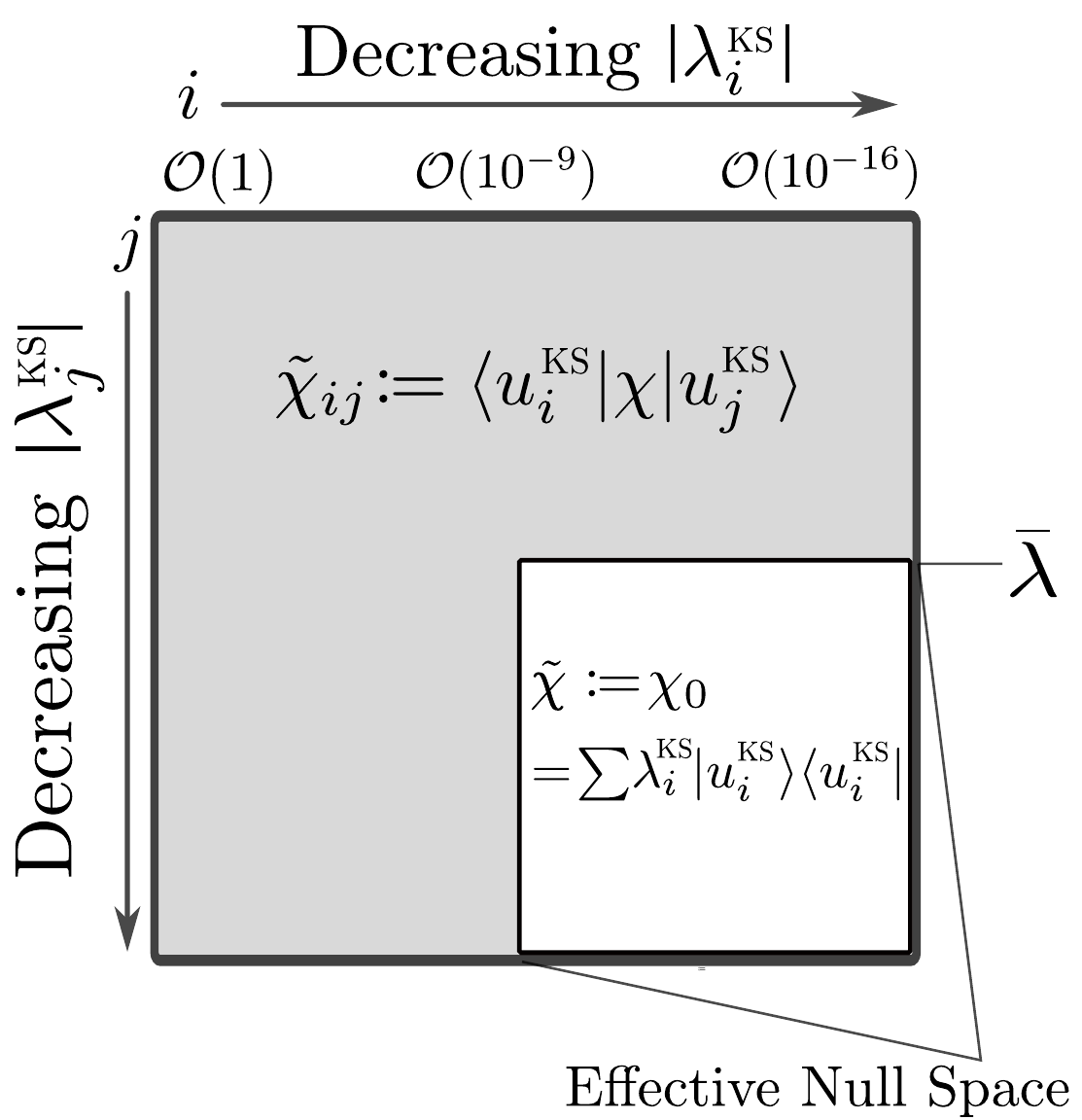}
\end{center}
\caption{A schematic depiction of the \textit{eigenspace truncation} strategy used to regularize computations of $f_\text{xc}$, whereby the interacting response function is expanded in the basis of eigenvectors of the Kohn-Sham response function, and set equal to the Kohn-Sham response function inside the effective null space, parameterized by $\bar{\lambda}$.}
\label{fig:eigenspaceTruncation}
\end{figure}

Consider the eigendecomposition of the interacting and Kohn-Sham response matrices at a given $\omega$, where the eigenpairs are denoted $\{|u_i \rangle, \lambda_i \}$ and $\{|u^\text{KS}_i \rangle, \lambda^\text{KS}_i \}$ respectively. Consider further some value $\bar{\lambda}$ such that the \textit{effective null space}, $\text{Null}_\text{eff}$, is defined as the subspace spanned by eigenvectors whose eigenvalue has modulus below $\bar{\lambda}$ \footnote{The formal definition of the effective null space is $\text{Null}_\text{eff}(\chi_0) = \text{Span}(\{ |u_i^\text{KS}\rangle \ | \ |\lambda^\text{KS}_i| < \bar{\lambda} \})$.}. The assumption is now made that the \textit{truncated response matrix} $\tilde{\chi}$ operates on vectors that are elements of the effective null space as the Kohn-Sham response matrix,
\begin{align}
\tilde{\chi} |v \rangle = \chi_0 |v\rangle \text{ for } |v\rangle \in \text{Null}_\text{eff}(\chi_0), \label{eq:eigentruncation}
\end{align}
see Fig$.$ \ref{fig:eigenspaceTruncation}. Given $P_N$ as the projection operator onto the effective null space, the expression in Eq$.$ (\ref{eq:eigentruncation}) is established as follows,
\begin{align}
\tilde{\chi} = (I - P_\text{N}) \chi + P_\text{N} \chi_0;
\end{align}
the first term on the right-hand side removes the effective null space from $\chi$, and the second term ensures $\tilde{\chi}$ operates as intended on elements of the effective null space. Another view of this manipulation is that the truncated and Kohn-Sham response functions are required to share eigenvectors and eigenvalues inside the effective null space,
\begin{align}
\{ |\tilde{u}_i\rangle, \tilde{\lambda}_i \} = \{ |u^{\text{KS}}_i\rangle, \lambda^{\text{KS}}_i \} \text{ for } |\lambda^\text{KS}_i| < \bar{\lambda},
\end{align}
and the truncated response function is otherwise equal to the interacting response function, this is also depicted in Fig$.$ \ref{fig:eigenspaceTruncation}.
The \textit{pseudoinverse} \citep{GenInv} with cutoff $\bar{\lambda}$, i.e. the eigendecomposition with eigenpairs below $\bar{\lambda}$ discarded, is now an exact procedure to obtain $f_\text{xc}$ that recovers $\tilde{\chi}$ after solution of the Dyson equation,
\begin{align}
f_\text{xc} = \chi_0^{+} - \tilde{\chi}^{+} - f_{\text{H}}, 
\end{align}
where $^{+}$ denotes the pseudoinverse. 

In direct analogy with the previous method, the parameter $\bar{\lambda}$ assumes the job of $b$, namely, it parameterizes the extent of the truncation. The difference between the truncated response function Eq$.$ (\ref{eq:eigentruncation}) and the exact interacting response function is again given the name \textit{method error}. Note that this approach is quite distinct from applying the pseudoinverse to the response functions in Eq$.$ (\ref{eq:xcKernel}) -- doing so would introduce much more error and the source of this error is not clear. On the contrary, the error inherent in the method presented here is identified as the extent to which the effective null space of the interacting and Kohn-Sham response functions do not overlap, and this error can be tracked without reference to $f_\text{xc}$ using the method error. 

Since the eigenspace truncation method is much more accurate than the real-space truncation method, results are given using the eigenspace truncation where possible. Although, visualization of the optical spectrum relies on evaluating the response functions slightly above the real axis in the frequency domain, along $\omega + i \eta$. This leads to response matrices that are complex-symmetric, and thus (weakly) non-Hermitian, in which case the real-space truncation method is used.

\subsection{Gauge freedom}
\label{sec:GaugeFreedomTheory}

As noted in Refs$.$ \citep{Hellgren2008,Hellgren2009,Hellgren2012,Aryasetiawan2002}, the following transformation
\begin{align}
f_\text{xc}(x,x',\omega) \rightarrow f_\text{xc}(x,x',\omega) + g(x,\omega) + h(x',\omega) + c(\omega), \nonumber
\end{align}
leaves the output of the Dyson equation unchanged, and thus we are, in principle, free to choose the arbitrary complex-valued functions $g(x,\omega), h(x',\omega)$ and $c(\omega)$. All three transformations are a direct manifestation of the invariance of quantum Hamiltonian systems under a constant time-dependent shift of the potential. From the point of view of $f_\text{xc}$ approximations, two xc kernels are equivalent if they exist within this family of functions \footnote{Note that the quantities $\langle ij | f_\text{xc}(\omega) | kl \rangle$, where $(i,j,k,l)$ label indices of single-particle wavefunctions, \textit{are} unique \citep{Hellgren2009}, and it is these quantities that form the input to the Casida equation \citep{Casida1995}, for example.}. 

A preferred gauge is defined by Eq$.$ (\ref{eq:xcKernel}), since the objects $\chi$ and $\chi_0$ are \textit{themselves} invariant to a shift in the potential. The unique $f_\text{xc}$, modulo a constant shift (see below), defined in Eq$.$ (\ref{eq:xcKernel}) can be considered the physical $f_\text{xc}$, and it is this definition of $f_\text{xc}$ that is assumed in discussions on its various properties and limits \citep{Ullrich2016,Botti2007,Ghosez1997,Byun2019,Godby1989}. To modify this $f_\text{xc}$ using its gauge freedom changes its underlying structure; for example, setting $g \neq 0$ gives $f_\text{xc}$ spurious long-range behavior, and setting $g \neq h$ produces an $f_\text{xc}$ that is not symmetric under interchange of $x \leftrightarrow x'$. In this work, we illustrate $f_\text{xc}$ as it is defined in Eq$.$ (\ref{eq:xcKernel}), which also \textit{defines} $g = h = 0$. The constant shift $c$ has special meaning, as it is itself an eigenvector of $\chi$ and $\chi_0$ with eigenvalue zero, i.e. the response functions are non-invertible in this direction. Since the Dyson equation is therefore silent regarding the value of $c$, we anchor $f_\text{xc}$ by requiring that, in the long-range limit, far outside the confined density, $f_\text{xc} + f_\text{H} \rightarrow 0$, and find that this limit is reliably achieved. 

Having decided upon a preferred gauge, one can consider the possible consequences of this gauge freedom on matters of practical interest. An approximate $f_\text{xc}$ that differs in relevant structure from the exact $f_\text{xc}$ largely due to a change of gauge provides at least a partial explanation for the performance of a given approximate $f_\text{xc}$; in Section \ref{sec:GaugeFreedom} we consider this line of inquiry.

\section{Results and discussion}
\label{sec:Results}

\subsection{Atom}

Our first system consists of two interacting electrons confined in the atom-like potential, $v_\text{ext}(x) = - 2 / (|0.1x| + 0.5)$, within the domain $[-8,8]$ a.u. An illustration of this system, and its associated numerical parameters, are given in the supplemental material. The purpose of the atom demonstration is two-fold. First, it constitutes a proof-of-concept, and defines a standard of accuracy to which the remainder of the calculations are held unless stated otherwise. Second, the optical spectrum of the atom is calculated in the range $\omega \in [0,6]$ a.u., which includes many excitations beyond the first, and the efficacy of various approximations to $f_\text{xc}$ are examined in relation to the optical spectrum. 

\begin{figure}[ht]
\begin{center}
\includegraphics[width=3.5in]{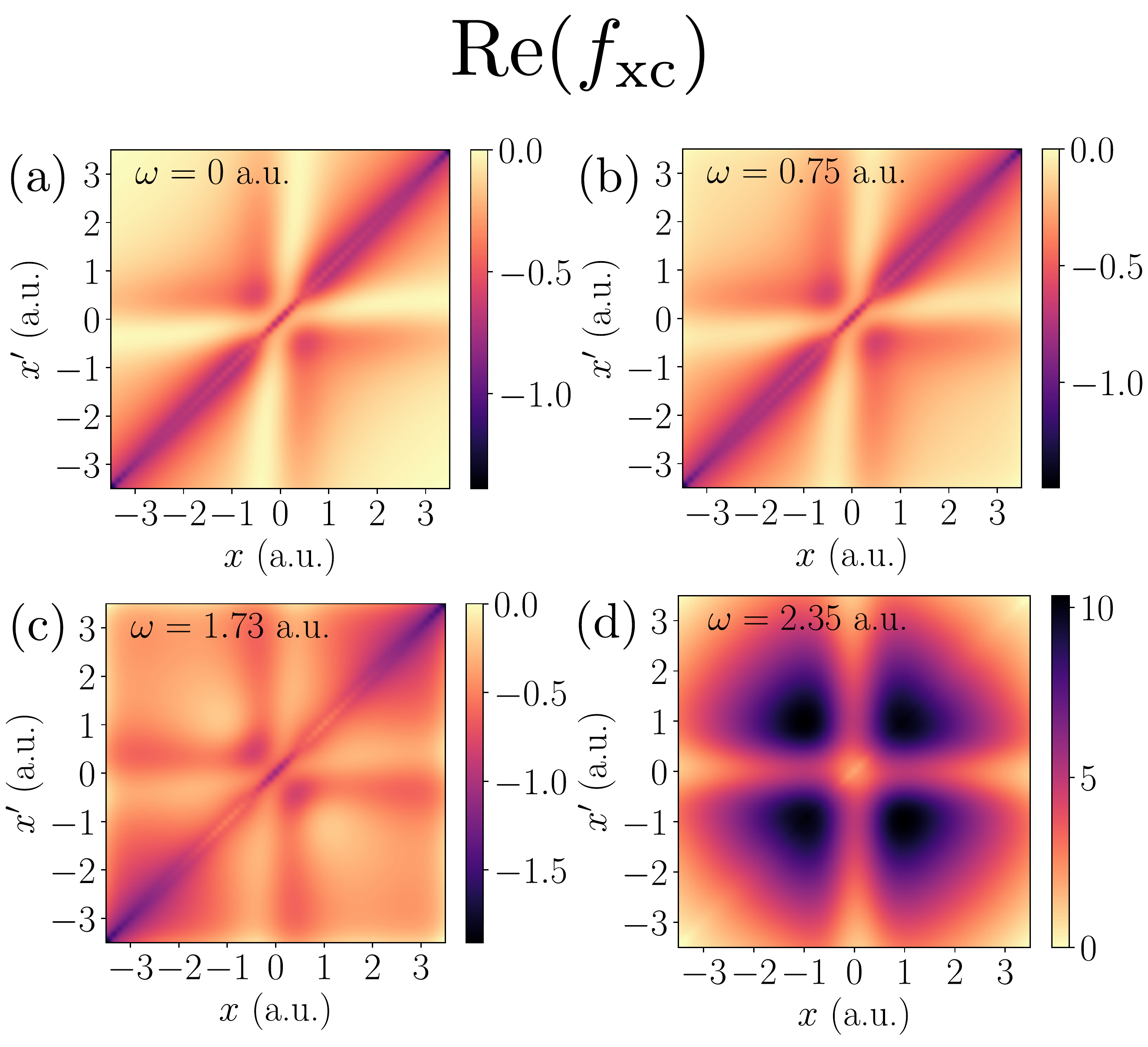}
\end{center}
\caption{The real part of the exact numerical xc kernel $f_\text{xc}(x,x',\omega)$ for the atom at (a) $\omega=0$ (exact adiabatic), (b) the first visible interacting excitation, (c) the second visible interacting excitation, and (d) the third visible interacting excitation. For illustrative purposes, $f_\text{xc}$ is shown for $-3.5 < x < 3.5$, where its essential structure is most visible. The xc kernel across the first two transitions remains approximately equal to the exact adiabatic $f_\text{xc}(\omega = 0)$, after which a significant departure from the adiabatic limit is observed. The exact adiabatic $f_\text{xc}(\omega = 0)$ displays considerable non-local structure, despite a local dominance along $x=x'$.}
\label{fig:pseudoatomicXCKernel}
\end{figure}

The exact $f_\text{xc}$, constructed using the real-space truncation method, is shown for the first three visible interacting excitations in the optical spectrum, and at $\omega = 0$ a.u., in Fig$.$ \ref{fig:pseudoatomicXCKernel}. The last is sometimes termed the \textit{exact adiabatic} $f_\text{xc}$, and it correctly describes any system in which the response to a perturbation is essentially instantaneous, $f_\text{xc}(x,x',\omega = 0)$. 

The real-space truncation parameter is chosen as $b=5.8$ a.u., meaning the inner region is defined as $-5.8 < x < 5.8$. It is important to stress that this choice of $b$ is \textit{not} unique, and there exists some feasible range of $b$ within which $f_\text{xc}$ itself is insensitive to changes. Moreover, within this feasible range, both the method error and numerical error are acceptable -- a discussion on the precise quantification of error here is given in the supplemental material. The mean absolute error between the output of the Dyson equation, which is hereafter defined as
\begin{align}
\chi_\text{Dyson}(\omega) = \frac{\chi_0(\omega)}{I - \chi_0(\omega) (f_\text{xc}(\omega) + f_\text{H})}, \label{eq:finiteBasisDyson}
\end{align}
and the interacting response function $\chi(\omega)$ is $\mathcal{O}(10^{-9})$ over the entire grid. The \textit{zero-force sum rule} \citep{Ullrich2012,Wagner2012} is used to further validate the numerics, which is discussed and illustrated in the supplemental material. 

\begin{figure}[ht]
\begin{center}
\includegraphics[width=3in]{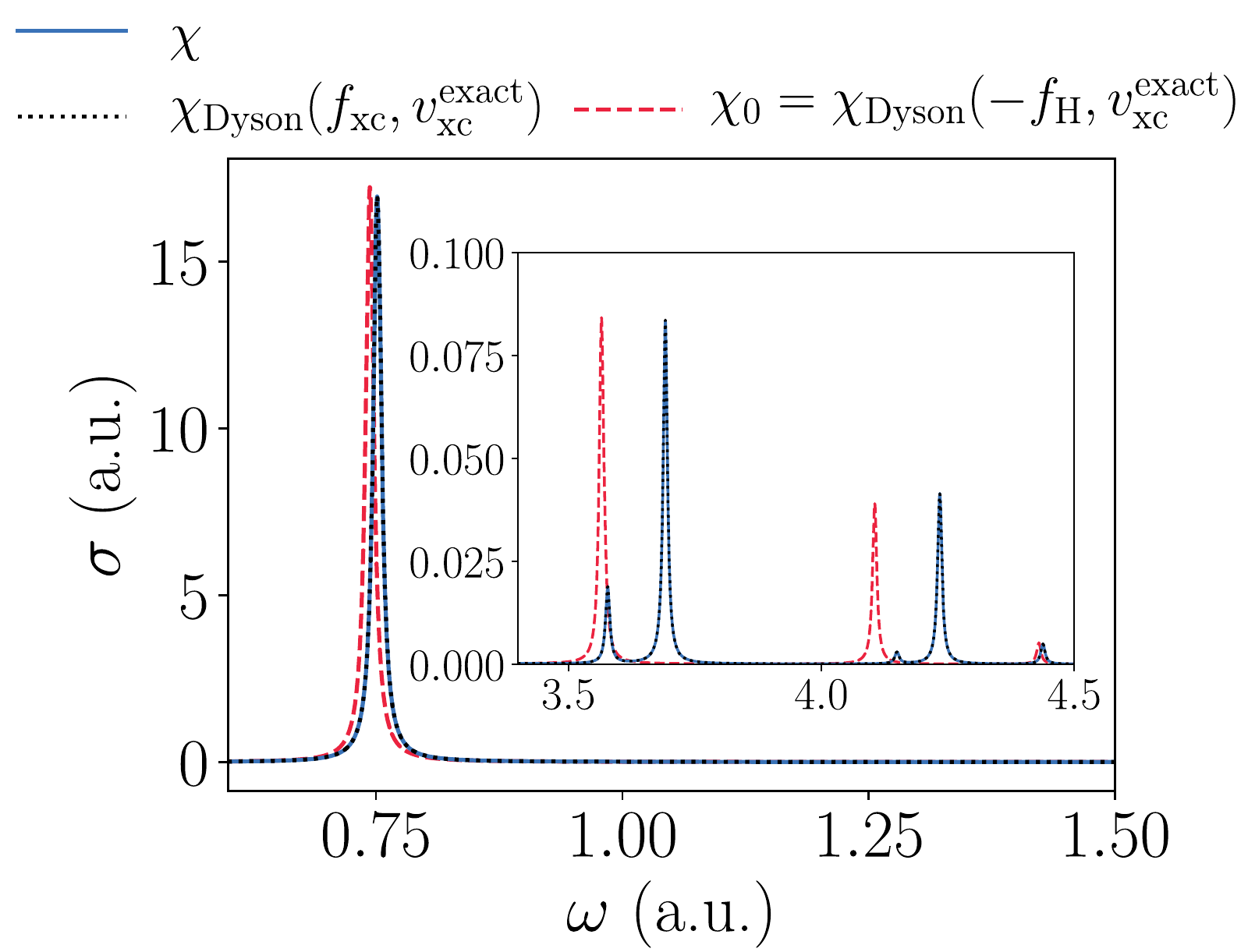}
\end{center}
\caption{The optical spectrum for the atom is computed using the interacting response function $\chi$ (blue solid), the Kohn-Sham response function $\chi_0$ (red dash), and the output of the Dyson equation $\chi_\text{Dyson}$ with the exact $f_\text{xc}(x,x',\omega)$ (black dot). The exact numerical $f_\text{xc}$ reproduces the interacting peaks perfectly, as expected.}
\label{fig:pseudoatomOAS_V1}
\end{figure}

In order to extract the optical transition energies and transition rates, given a single-particle response function $\chi_0$ and xc kernel $f_\text{xc}$, one can construct the entire optical absorption spectrum Eq$.$ (\ref{eq:opticalSpectrum}) using the corresponding output of the Dyson equation, denoted $\chi_\text{Dyson}(f_\text{xc}, v_\text{xc})$, where $\chi_0$ is specified with some $v_\text{xc}$, see Fig$.$ \ref{fig:pseudoatomOAS_V1}.

As established in \citep{Appel2003,Savin1998,AlSharif1998}, the exact Kohn-Sham single-particle transitions are in excellent agreement with the interacting transitions, but this agreement becomes increasingly poor at higher energies. An approach to understanding this is to consider the overlap between the final states involved in a given interacting $|\Psi_0\rangle \rightarrow |\Psi_f\rangle$ and Kohn-Sham $|\Phi_{(0,1)} \rangle \rightarrow |\Phi_f\rangle$ transition, where $|\Phi_{(i,j)}\rangle$ denotes the Slater determinant constructed from $i^\text{th}$ and $j^\text{th}$ Kohn-Sham single-particle states.

The overlap of the ground-state is $\langle \Psi_0 | \Phi_{(0,1)} \rangle = 0.99991$, and the overlap of the final states involved in the first transition at $\omega = 0.76$ a.u. is $\langle \Psi_1 | \Phi_{(0,2)} \rangle = 0.9995$. The \textit{static correlation} in the interacting state here is modest, meaning the interacting state has strong single-particle character, which leads to agreement between the low-energy transitions in the optical spectrum. At higher energies, the overlap decays by multiple orders of magnitude, however this is not the predominant source of error in higher energy transitions. Rather, interacting excitations that correspond to Kohn-Sham single-particle excitations out of the highest occupied state (the second)  are much more accurate than single-particle excitations out of the first state. For example, the interacting excitation $|\Psi_0\rangle \rightarrow |\Psi_{19}\rangle$ at $\omega = 4.41$ a.u. in Fig$.$ \ref{fig:pseudoatomOAS_V1} is captured well with the Kohn-Sham excitation $|\Phi_{(0,1)}\rangle \rightarrow |\Phi_{(0,12)}\rangle$, whereas this is not true of the preceding interacting excitation. This is not surprising, as the highest occupied Kohn-Sham state has energy equal to minus the \textit{exact} electron removal energy \citep{Perdew1982}, and thus at least one energy involved in the transition is correct. This might often be the case, and to comment further would require additional many-body calculations.

\begin{figure}[ht]
\begin{center}
\includegraphics[width=3.4in]{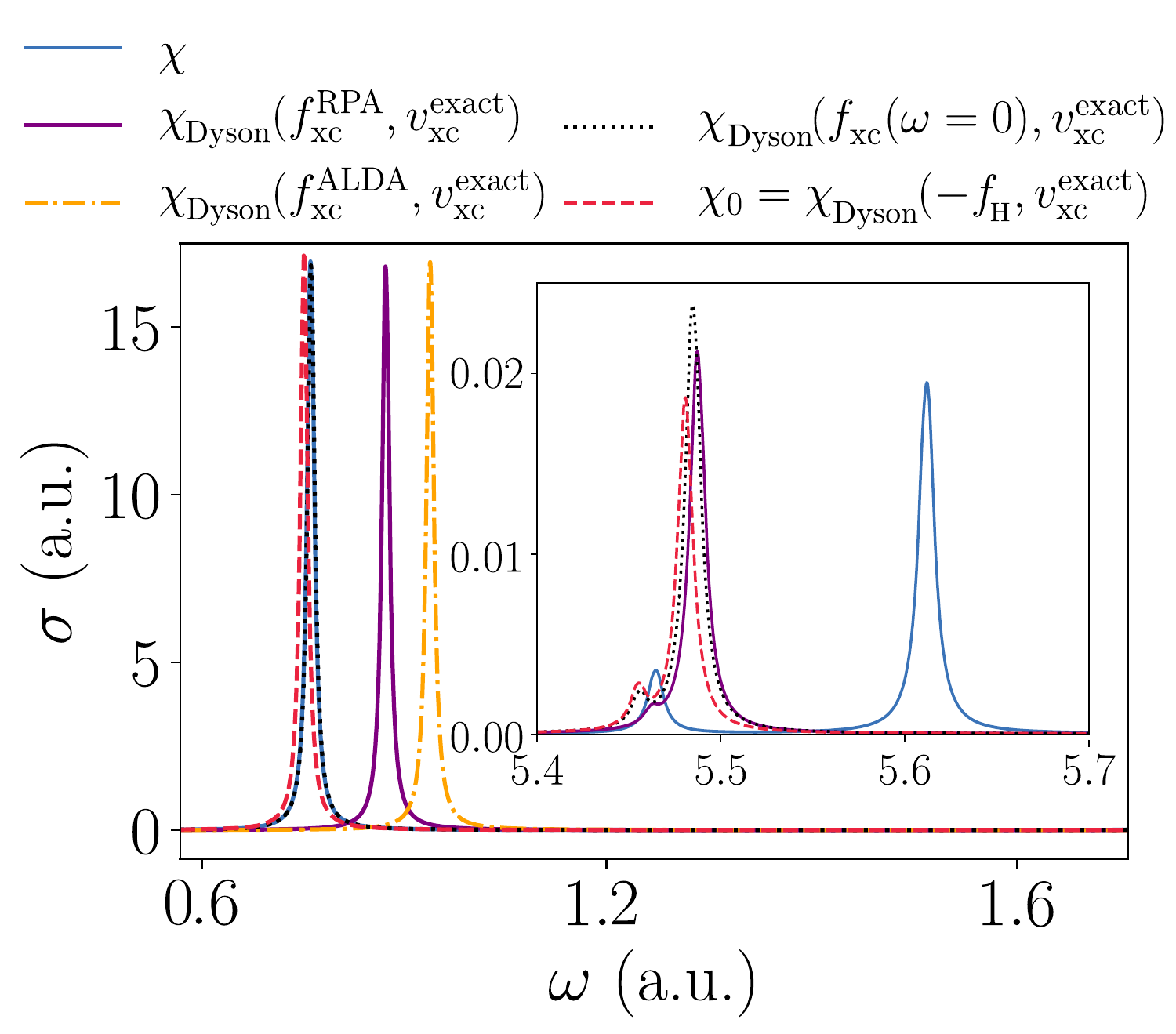}
\end{center}
\caption{The optical absorption spectrum for the atom around the first transition, and around two higher energy transitions (inset), calculated at various levels of approximation. The optical spectra calculated from the interacting and Kohn-Sham response functions are plotted alongside the optical spectrum computed from the Dyson equation using the RPA, adiabatic LDA, and exact adiabatic xc kernels. The adiabatic LDA and RPA fail to accurately reproduce the interacting transitions, whereas the exact adiabatic approximation successfully describes the low-energy transition, but fails at higher energies.}
\label{fig:pseudoatomOAS_V2}
\end{figure}

The following $f_\text{xc}$ approximations are now considered: the RPA $f^\text{RPA}_\text{xc} = 0$, an adiabatic LDA $f^\text{ALDA}_\text{xc}[n](x,x',\omega=0) \propto \delta(x-x')$ parameterized with reference to the homogeneous electron gas in \citep{Entwistle2018}, and the exact adiabatic xc kernel $f_\text{xc}(x,x',\omega=0)$, Fig$.$ \ref{fig:pseudoatomicXCKernel}(a). These $f_\text{xc}$ approximations are used to solve the Dyson equation in conjunction with the \textit{exact Kohn-Sham response function}. The atomic optical spectrum, using the aforementioned series of approximations, is shown in Fig$.$ \ref{fig:pseudoatomOAS_V2} for the first transition and a chosen higher energy transition.

The exact adiabatic $f_\text{xc}$ reproduces the first peak well, which reflects the fact that $f_\text{xc}$ at the first excitation, Fig$.$ \ref{fig:pseudoatomicXCKernel}(b), is visually indistinguishable from the exact adiabatic $f_\text{xc}$. This agreement demonstrates that, not only is the adiabatic approximation valid for low-energy transitions, but also the non-local spatial structure in the exact adiabatic $f_\text{xc}$ is required in order to reproduce the low-energy peaks in the optical spectrum -- in lacking this structure, both the RPA and adiabatic LDA significantly over-correct the underestimation of the non-interacting transition energy. At higher energies, i.e. beyond the third peak in the optical spectrum (not shown), all three approximations perform similarly, and do not improve matters significantly beyond the corresponding non-interacting peak. This behavior appears to be generic for all peaks observed up to $\omega = 6$ a.u., namely, the transition energies output from the Dyson equation with these approximate xc kernels are bound to the quality of the non-interacting transition energies. Furthermore, the inset of Fig$.$ \ref{fig:pseudoatomOAS_V2} demonstrates that the exact adiabatic approximation gives an excitation energy that is worse than the other \textit{adiabatic} approximations. This suggests that, in order to capture higher energy transitions, a frequency dependence is required, and in particular `improving' the spatial structure of adiabatic approximations toward the exact adiabatic structure does not assist matters here.  

As alluded to above, the exact adiabatic approximation ceases to out-perform the adiabatic LDA and RPA beyond the third transition in the optical spectrum, i.e. the same transition for which the corresponding \textit{exact} $f_\text{xc}$ departs from its adiabatic structure in a serious manner, Fig$.$ \ref{fig:pseudoatomicXCKernel}. In fact, as the subsection to follow demonstrates, this violent departure from adiabaticity has a particular and fairly simple form, and its origin is understood in the context of eigenvalues of $\chi(\omega)$ or $\chi_0(\omega)$ that \textit{cross} zero. The eigenvector corresponding to the eigenvalue that touches zero is a non-$v$-representable density perturbation at linear order.

\subsection{Infinite potential well}
\label{sec:InfPotWell}

The infinite potential well is defined with the external potential $v_\text{ext} = 0$ inside the domain $[-5,5]$ a.u. (See supplemental material for the corresponding Kohn-Sham potential, density, and numerical parameters.) The numerical response functions for this system are well-conditioned and valid up to arbitrary $\omega$, there are no regions of nearly vanishing density, and thus there is no significant numerical error present here. 

\begin{figure}[ht]
\begin{center}
\includegraphics[width=3.3in]{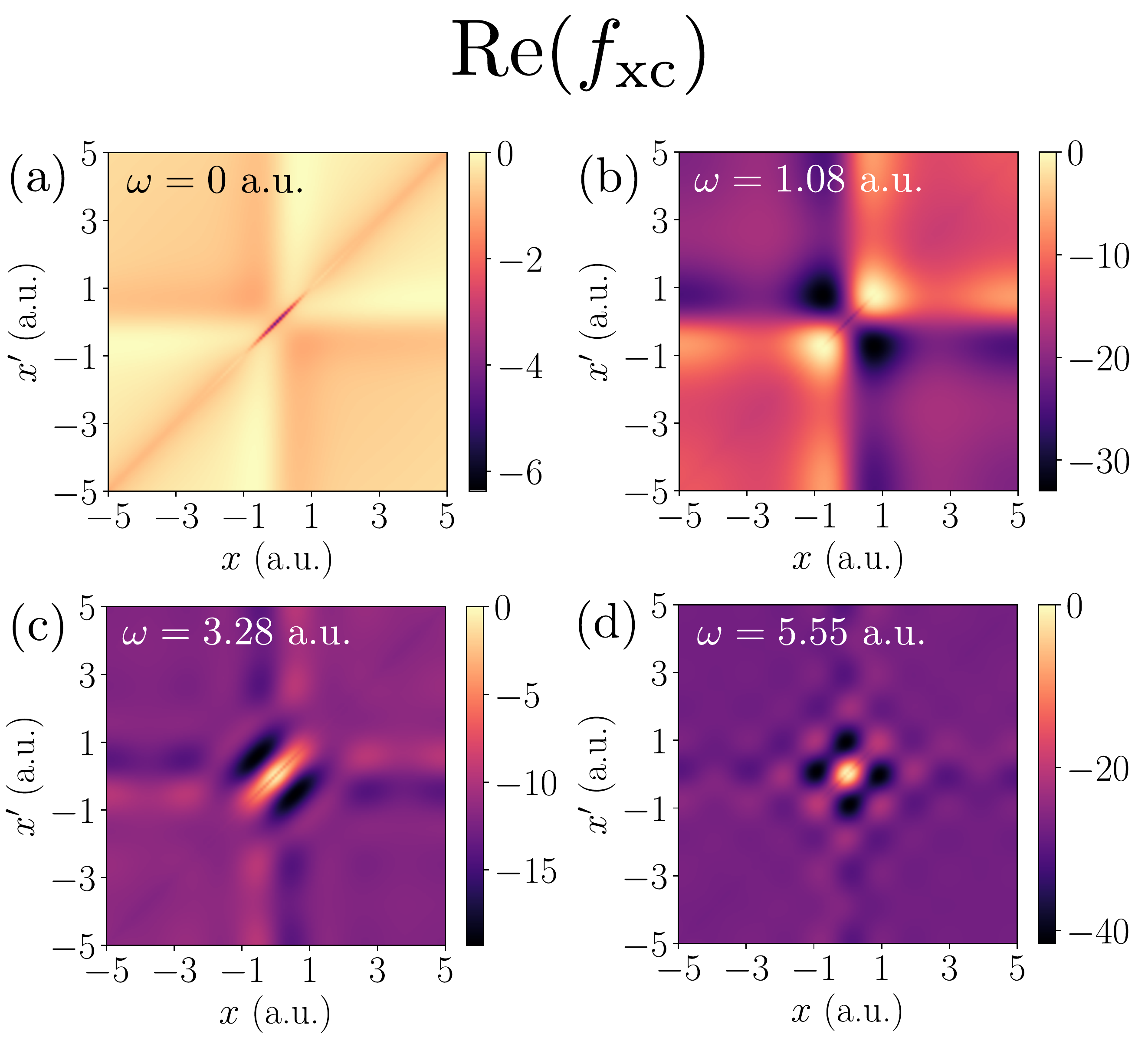}
\end{center}
\caption{The real part of the exact numerical xc kernel $f_\text{xc}(x,x',\omega)$ for the infinite potential well at (a) $\omega=0$ (exact adiabatic), and at three higher frequencies that demonstrate its non-adiabatic character, (b) the sixth interacting excitation, $\omega = 1.09$ a.u., (c) $\omega = 3.28$ a.u., and (d) $\omega = 5.55$ a.u. The xc kernels are shifted such that their maximum value is zero, since there exists no long-range limit, see Section \ref{sec:GaugeFreedomTheory}. A strong frequency dependence, i.e. departure from the adiabatic limit, is observed.}
\label{fig:infPotWellfxc}
\end{figure}

\begin{figure}[ht]
\begin{center}
\includegraphics[width=3.3in]{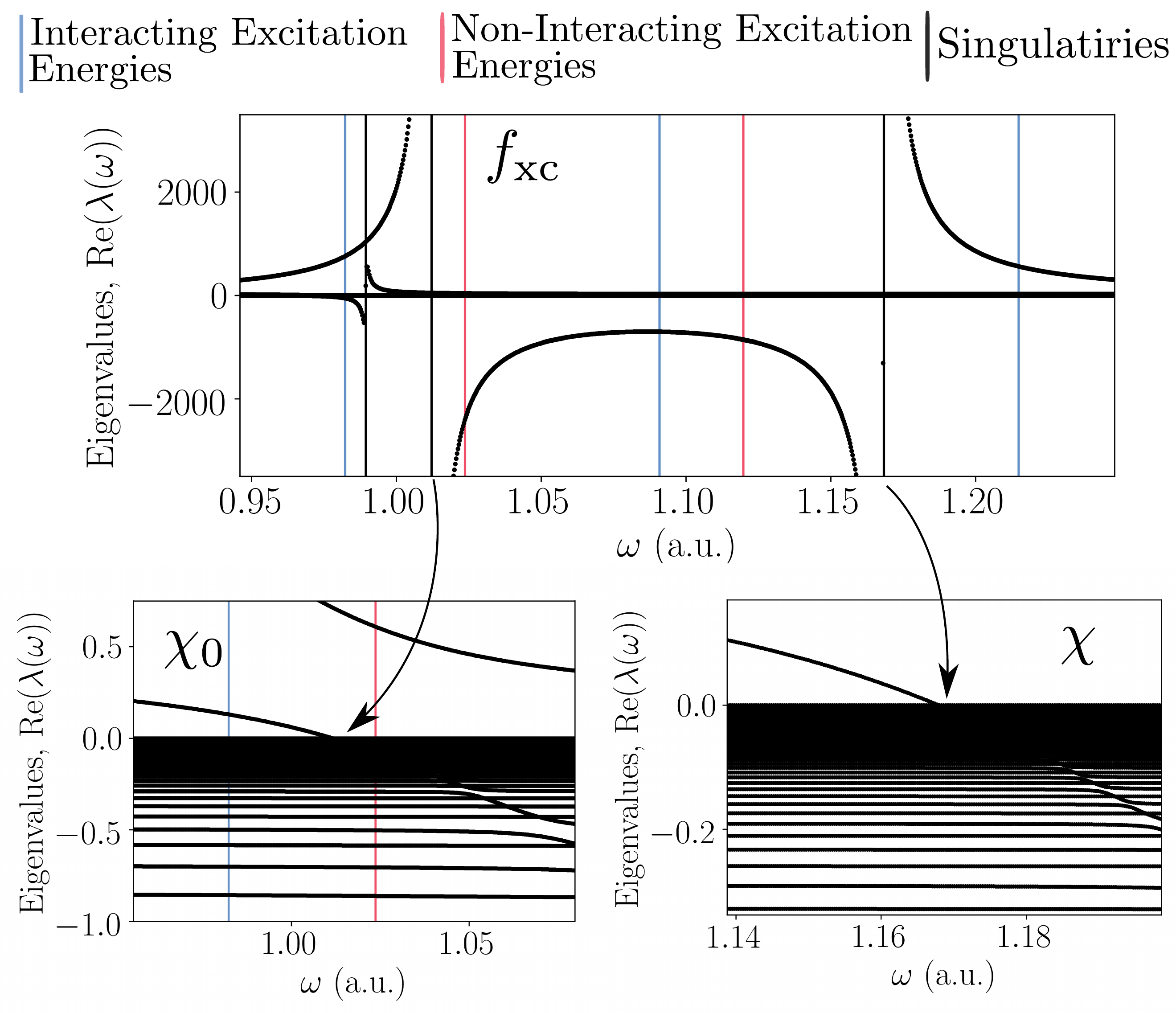}
\end{center}
\caption{Eigenvalues as a function of frequency, labeled Re($\lambda(\omega)$), of $f_\text{xc}$ (upper), $\chi_0$ (lower left), and $\chi$ (lower right). Within the frequency range shown, the predominant non-adiabatic behavior of $f_\text{xc}$ is a result of three singularities and their surrounding divergences at $\omega = 0.99, 1.01, 1.17$ a.u. The source of the last two is observed to be an eigenvalue crossing zero in $\chi_0$ and $\chi$ respectively.}
\label{fig:infPotWellEigenvalues}
\end{figure}

The non-adiabatic character of $f_\text{xc}$ is illustrated up to $\omega \approx 6$ a.u. in Fig$.$ \ref{fig:infPotWellfxc}. As in the atom, $f_\text{xc}$ exhibits little frequency dependence, until after the second transition. This behavior is related to the \textit{poles} that occur in $f_\text{xc}$ infinitesimally below the $\omega$-axis \citep{VanLeeuwen2001, Hellgren2009}. The eigenvalues of $f_\text{xc}$ as a function of frequency are given in Fig$.$ \ref{fig:infPotWellEigenvalues}, and in particular, three divergences are shown (the singularities are tempered slightly by evaluating the response functions just above the real $\omega$-axis). The lower panels of Fig$.$ \ref{fig:infPotWellEigenvalues} demonstrate that these singularities coincide with a single eigenvalue of either the interacting or non-interacting response function crossing zero. Thus, the visual character of $f_\text{xc}$ is dominated by the outer product of the eigenvector whose eigenvalue is either beginning to diverge, or recovering from a divergence. Moreover, nothing in principle is preventing these singularities in $f_\text{xc}$ coming arbitrarily close to an interacting excitation -- either $\chi_0$ can cross zero close to an interacting excitation, or $\chi$ itself can cross zero close to an interacting excitation \footnote{The interacting response function $\chi$ diverges at an interacting excitation, but this does not, in a finite basis, prevent an eigenvalue from crossing zero at this energy under certain circumstances that are set out in the supplemental material.}. Since most $f_\text{xc}$ approximations lack these divergences, one can question the importance of these divergences in relation to optical properties.

In order to determine the impact of the divergences in $f_\text{xc}$ on the optical spectrum, we examine how the optical spectrum is affected after projecting out the divergent eigendirection in $f_\text{xc}$, but otherwise keeping $f_\text{xc}$ identical (the details of this procedure are given in the supplemental material). This is tantamount to setting $\chi |v\rangle \coloneqq \chi_0|v\rangle$ for some vector $|v\rangle$ within the Hilbert space that is associated with the divergence. 

\begin{figure}[ht]
\begin{center}
\includegraphics[width=3.3in]{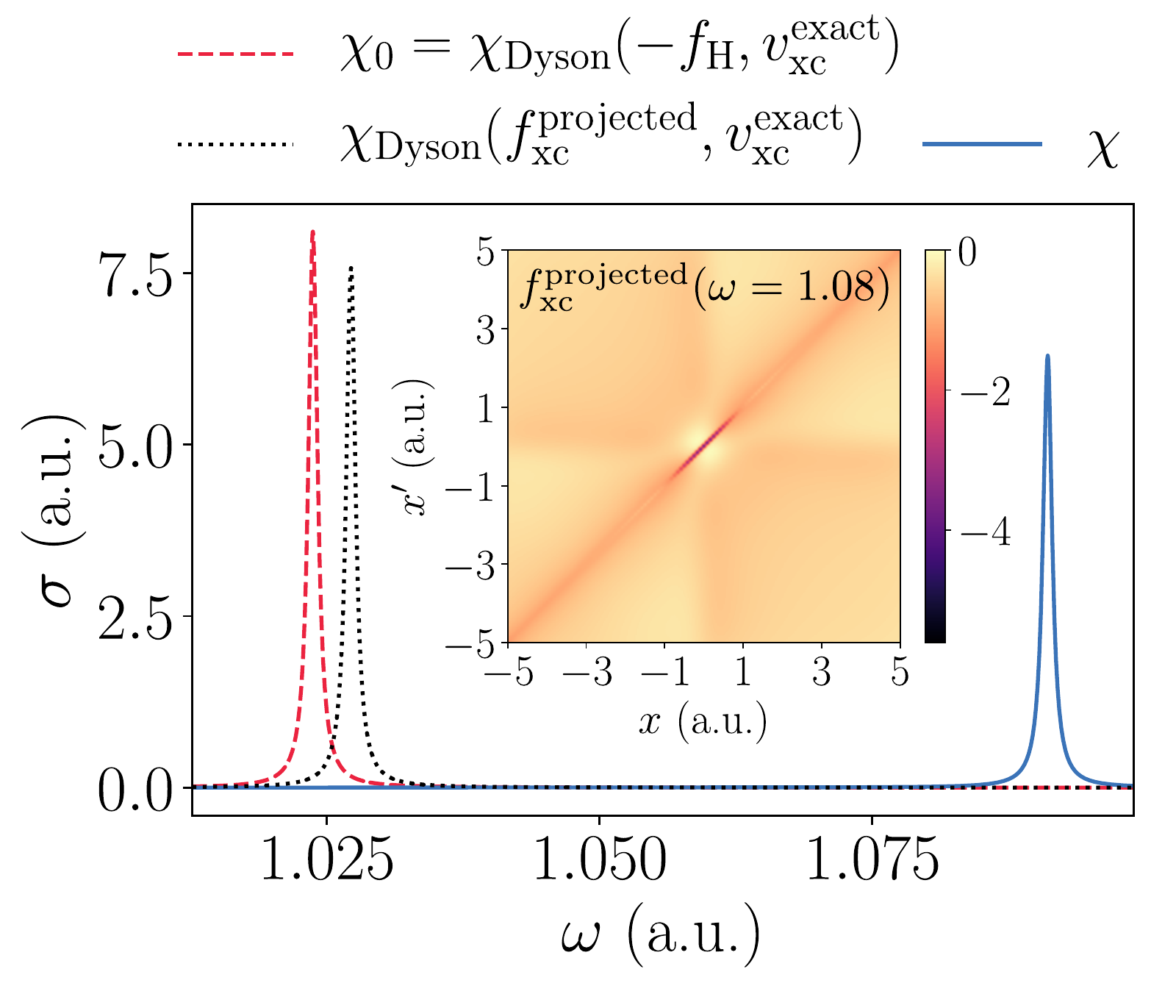}
\end{center}
\caption{The optical absorption spectrum for the infinite potential well around a visible interacting excitation at $\omega = 1.094$ a.u, and its corresponding non-interacting excitation at $\omega = 1.024$ a.u. The projected xc kernel (inset), i.e. $f_\text{xc}$ with its divergences removed, is indistinguishable from the adiabatic xc kernel, see Fig$.$ \ref{fig:infPotWellfxc}(a), and the optical peak associated with $f_\text{xc}^\text{projected}$ is only a slight improvement on the non-interacting peak.}
\label{fig:infPotWellOASProjected}
\end{figure}

The interacting excitation at $\omega = 1.09$ a.u., see Fig$.$ \ref{fig:infPotWellEigenvalues}, is visible in the optical spectrum, and furthermore the character of $f_\text{xc}$ at this energy, see Fig$.$ \ref{fig:infPotWellfxc}, is dominated by the outer product of an eigenvector whose eigenvalue is much larger in magnitude than the rest, and between the two divergences at $\omega = 1.01,1.17$ a.u. The removal of these divergences from $f_\text{xc}$ across the energy range of interest, and subsequently solving the Dyson equation with the \textit{projected xc kernel} $f_\text{xc}^\text{projected}$, shifts the interacting optical peak back toward the non-interacting peak, Fig$.$ \ref{fig:infPotWellOASProjected}. Conversely, the much weaker divergence at $\omega = 0.99$ a.u., as seen in Fig$.$ \ref{fig:infPotWellEigenvalues}, has a tail that also yields an eigenvalue much larger in magnitude than the rest at the interacting excitation, and removal of this divergence produces no change in the optical spectrum, i.e. this eigenvector is \textit{not} relevant for capturing the transition in question. In the inset of Fig$.$ \ref{fig:infPotWellOASProjected}, the visual character of $f_\text{xc}$ is shown at the interacting excitation $\omega = 1.09$ a.u. after the divergences visible in Fig$.$ \ref{fig:infPotWellEigenvalues} have been removed. Underneath these divergences, $f_\text{xc}$ is indistinguishable from the adiabatic $f_\text{xc}$, meaning the frequency dependence of $f_\text{xc}$ in this system is largely due to its pole structure. These results suggest that functional approximations that do not capture the non-adiabatic pole structure of $f_\text{xc}$ will struggle to improve matters beyond the non-interacting peaks for certain transitions. 

A further point of note is that there are many more zeros in the interacting response function than the non-interacting response function, as a simple counting argument is sufficient to demonstrate. All $N$ eigenvalues of the response functions begin negative \citep{VanLeeuwen2001}, and each excitation brings a negative eigenvalue to a positive eigenvalue across a divergence, which otherwise evolves as a continuous function of $\omega$. For two electrons discretized with a basis set of dimension $N$, there are $\frac{1}{2}N(N-1) - 1$ interacting excitations, and $2N - 4$ non-interacting excitations, the difference being made up of double (triple, etc. with more than two electrons) excitations that are notoriously not present in the Kohn-Sham response function \citep{Maitra2004}. Therefore, there are a great deal more eigenvalues that must pass through zero in the interacting response function, and moreover, these eigenvalues that cross zero are connected to the excitations that require them to do so -- an account of the precise conditions under which this occurs is given in the supplemental material using a two-state model, see also \citep{Entwistle2018}.

In Fig$.$ \ref{fig:infPotWellEigenvalues}, there are three divergences, two of which are \textit{paired} at $\omega = 1.01, 1.17$ a.u., meaning the zero in the non-interacting response function is a shifted version of the zero in the interacting response function; such behavior can be related to single excitations. There also exists an \textit{unpaired} divergence at $\omega = 0.99$ a.u. related to the excitation at $\omega = 0.98$ a.u. which has \textit{double excitation character}. That is, the overlap between the final state involved in this transition $|\Psi_0\rangle \rightarrow |\Psi_5\rangle$ and the Slater determinant $|\Phi_{(2,3)} \rangle$ is $\langle \Phi_{(2,3)} | \Psi_5\rangle = 0.98$.  

It is perhaps, then, no surprise that removal of the eigenvector and eigenvalue whose source is the unpaired divergence did not alter the visible (single excitation) transition in Fig$.$ \ref{fig:infPotWellOASProjected}. In fact, the pole in the interacting response function relating to the double excitation \textit{disappears} with removal of the unpaired divergence, and so this divergence, and its surrounding character, \textit{is} important in order to capture the transition for which it is relevant. The xc kernel exhibits unpaired divergences after all double excitations up to $\omega = 6$ a.u. at frequencies slightly higher than the interacting double excitation energy. Refs. \citep{Maitra2004,Cave2004} derive, under certain assumptions, the necessarily divergent character of $f_\text{xc}$ around a double excitation. The above analysis reveals that divergences in $f_\text{xc}$ are, in fact, common and associated with the $\omega$ neighborhood containing multiple and single excitations alike.

\subsection{Quantum harmonic oscillator}

The quantum harmonic oscillator is defined in the domain $[-8,8]$ a.u. with the potential $v_\text{ext}(x) = \frac{1}{2} \nu^2 x^2$ where $\nu \coloneqq 0.45$ a.u. (See supplemental material for the corresponding exact Kohn-Sham potential, the interacting ground-state density, and the numerical parameters.) 

First, the spatial structure, and in particular the \textit{long-range behavior} \citep{Ullrich2016}, of the exact $f_\text{xc}$ is examined. The delicate nature of the numerics involved in computing the exact $f_\text{xc}$ is brought to the fore when attempting to capture its long-range limit. The ill-conditioning in the response matrices can be identified with regions of nearly vanishing ground-state density, see Section \ref{sec:BackgroundNumericalChallenges}, and it is precisely in these regions that we expect to observe the long-range character of $f_\text{xc}$. However, $f_\text{xc}$, when evaluated outside some central region where we can be confident there is little-to-no numerical error, diverges in a manner that is not consistent with the known long-range limit of $f_\text{xc}$. This spurious divergence in $f_\text{xc}$ does \textit{not} much affect the accuracy of the output of the Dyson equation, $\chi_\text{Dyson}$, because the Dyson equation Eq$.$ (\ref{eq:finiteBasisDyson}) involves the matrix product $\chi_0 f_\text{xc}$, and the divergent regions of $f_\text{xc}$ operate on the nearly vanishing regions of $\chi_0$. 

\begin{figure}[ht]
\begin{center}
\includegraphics[width=2.5in]{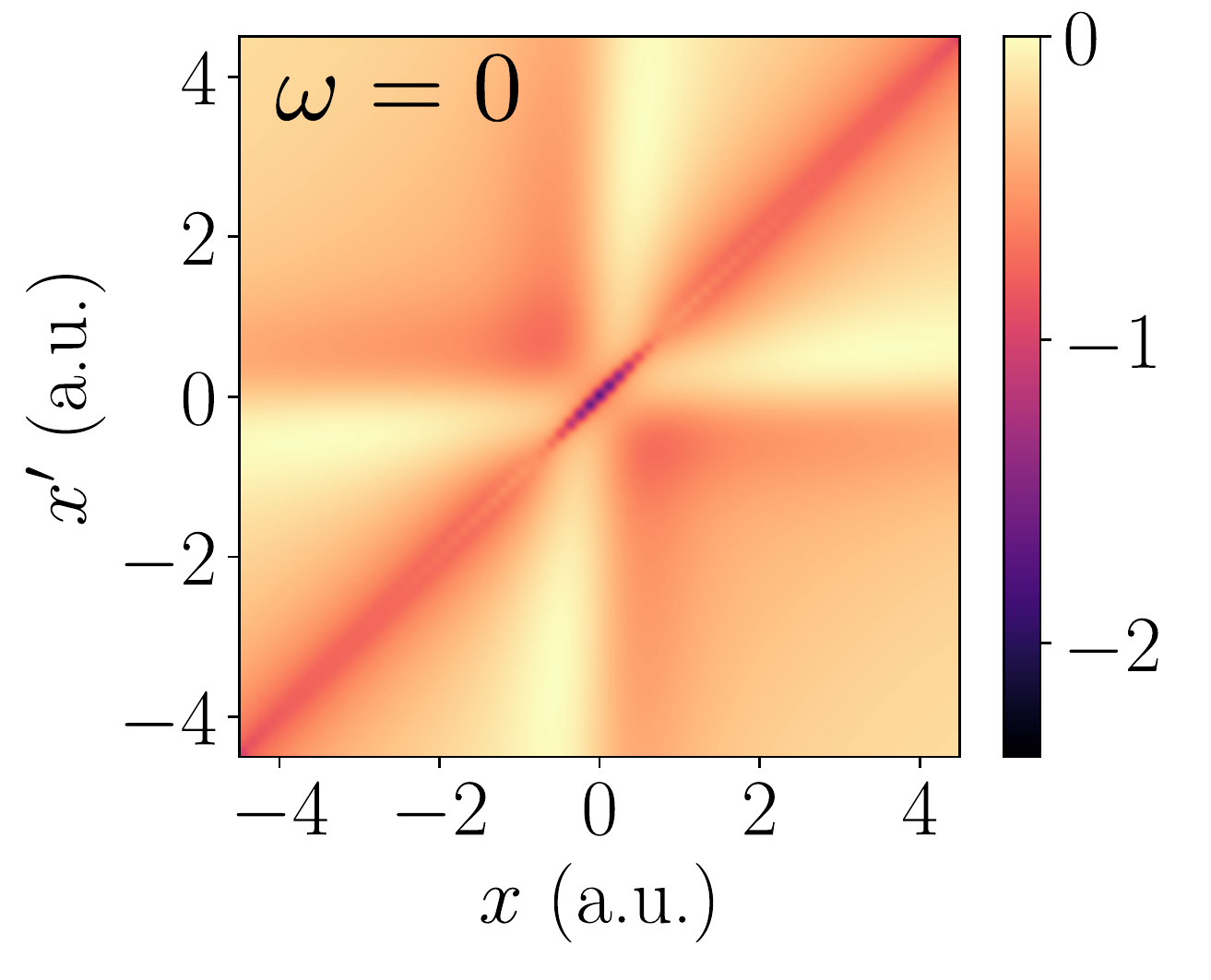}
\end{center}
\caption{The exact adiabatic xc kernel $f_\text{xc}(x,x',\omega=0)$ for the quantum harmonic oscillator constructed with the eigenspace truncation method. The precise spatial structure exhibited by the exact adiabatic kernel, including its long-range limit and non-local character, is discussed in the main text.}
\label{fig:QHOExactAdiabatic}
\end{figure}

\begin{figure}[ht]
\begin{center}
\includegraphics[width=3in]{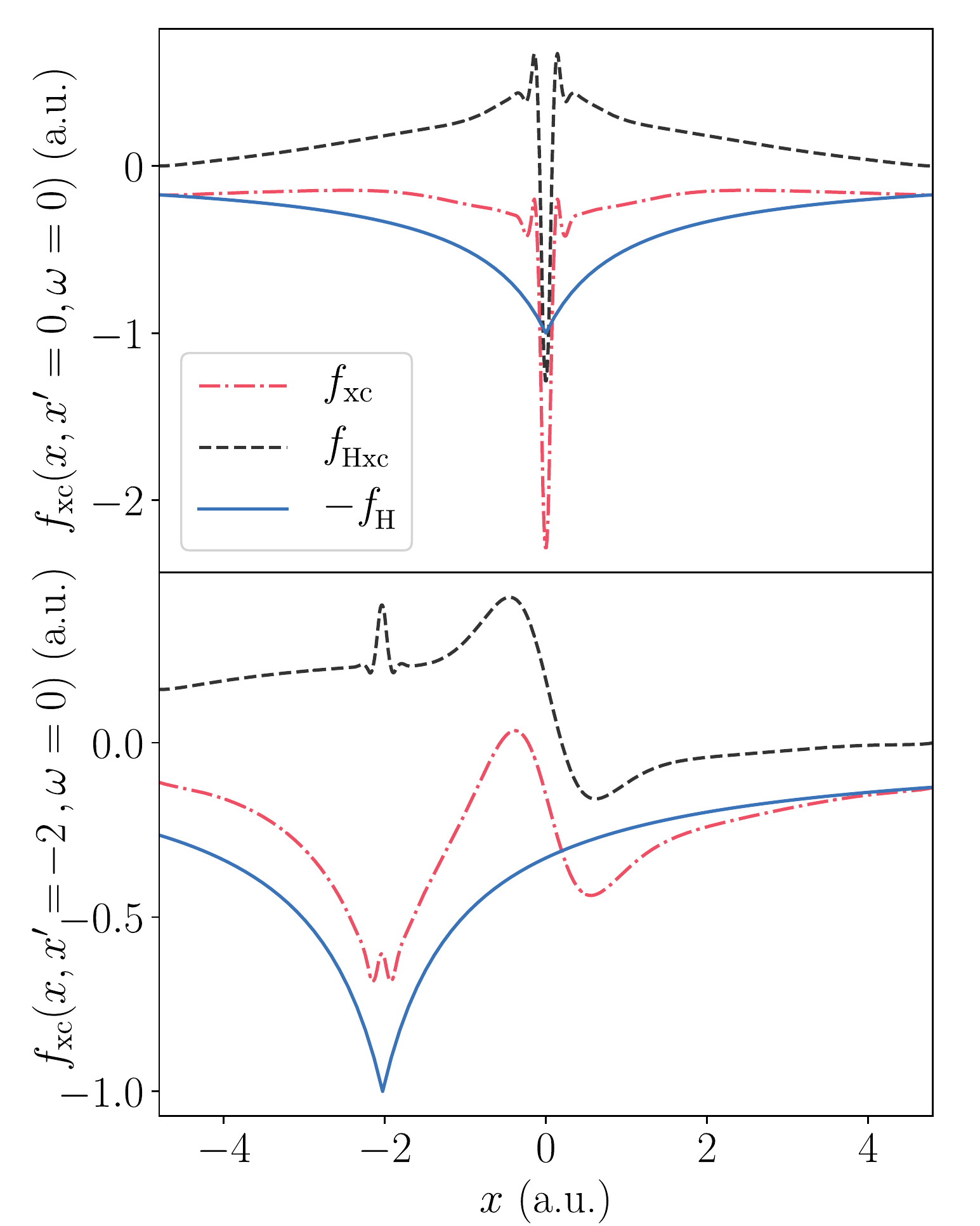}
\end{center}
\caption{The exact adiabatic xc kernel $f_\text{xc}(x,x',\omega=0)$ for the quantum harmonic oscillator along $x'=0$ a.u. (top) and $x'=-2$ a.u. (bottom). The xc kernel, Hxc kernel, and Hartree kernel, are shown, and the long-range limit $f_\text{xc} \rightarrow - f_\text{H}$ is observed.}
\label{fig:longRangeLimitQHOV2}
\end{figure}

If we are to observe the long-range limit of $f_\text{xc}$ in the present context, the region where the numerical error is low must overlap with the region where the long-range limit is observed. This is the case for the exact adiabatic $f_\text{xc}$, which is shown, together with slices of $f_\text{xc}$ along a particular axis, in Fig$.$ \ref{fig:QHOExactAdiabatic} and Fig$.$ \ref{fig:longRangeLimitQHOV2}.  

The center-most region of the domain is where exchange and correlation effects are most important, and in this region $f_\text{xc}$ exhibits a strong \textit{local} response that quickly decays as $x \neq x'$. Since this region can be interpreted as the most crucial for recovering accurate observable properties from the Dyson equation, this observation supports, at least in part, local approximations to $f_\text{xc}$, such as the adiabatic LDA. 

The non-local structure of $f_\text{xc}$ at $\omega=0$ a.u. can be seen most clearly in the lower panel of Fig$.$ \ref{fig:longRangeLimitQHOV2}, where perturbations in the density \textit{outside} the center-most region cause a significant change in the exchange-correlation potential \textit{inside} the center-most region. The failure of the adiabatic LDA to capture this non-local response leads to fairly poor agreement between the exact and approximate transition energies, which is seen for the atom in Fig$.$ \ref{fig:pseudoatomOAS_V2}, and illustrated for the quantum harmonic oscillator below \footnote{It is possible that the particular form of the non-local $f_\text{xc}$ functionals considered in \citep{Olsen2014,Olsen2019}, in which $f_\text{xc}(x,x') = f_\text{xc}[(n(x) + n(x'))/2]$, can capture certain features of the adiabatic non-local response depicted in  Fig$.$ \ref{fig:QHOExactAdiabatic} and Fig$.$ \ref{fig:longRangeLimitQHOV2}.}.

The long-range limit $f_\text{xc}(x,x',\omega=0) \rightarrow -f_\text{H}(|x-x'|)$ is explored in Fig$.$ \ref{fig:longRangeLimitQHOV2} along both $x'=0$ a.u. and $x'=-2$ a.u. Due to the rapid decay of the ground-state density in the quantum harmonic oscillator, convergence of $f_\text{xc}$ toward $-f_\text{H}$ along $x=-2$ a.u. is not directly observed in the $x \rightarrow -\infty$ limit -- the atom, whose ground-state density does not decay as quickly, converges toward $-f_\text{H}$ in both the positive and negative limits (see supplemental material). It is known  that the long-range limit of $f_\text{xc}$, in both finite and periodic systems, satisfies $f_\text{xc}(\omega) \rightarrow -\alpha(\omega) f_\text{H}$ \citep{Ullrich2016,Botti2007,Ghosez1997,Byun2019}, where $\alpha(\omega)$ is a frequency-dependent constant that reflects dielectric screening in the system. Therefore, one expects $\alpha=1$ in finite systems, and at lower frequencies, where the numerical methodology provides a robust long-range limit, our observations are consistent with this, see Fig$.$ \ref{fig:longRangeLimitQHOV2}.

\begin{figure}[ht]
\begin{center}
\includegraphics[width=3in]{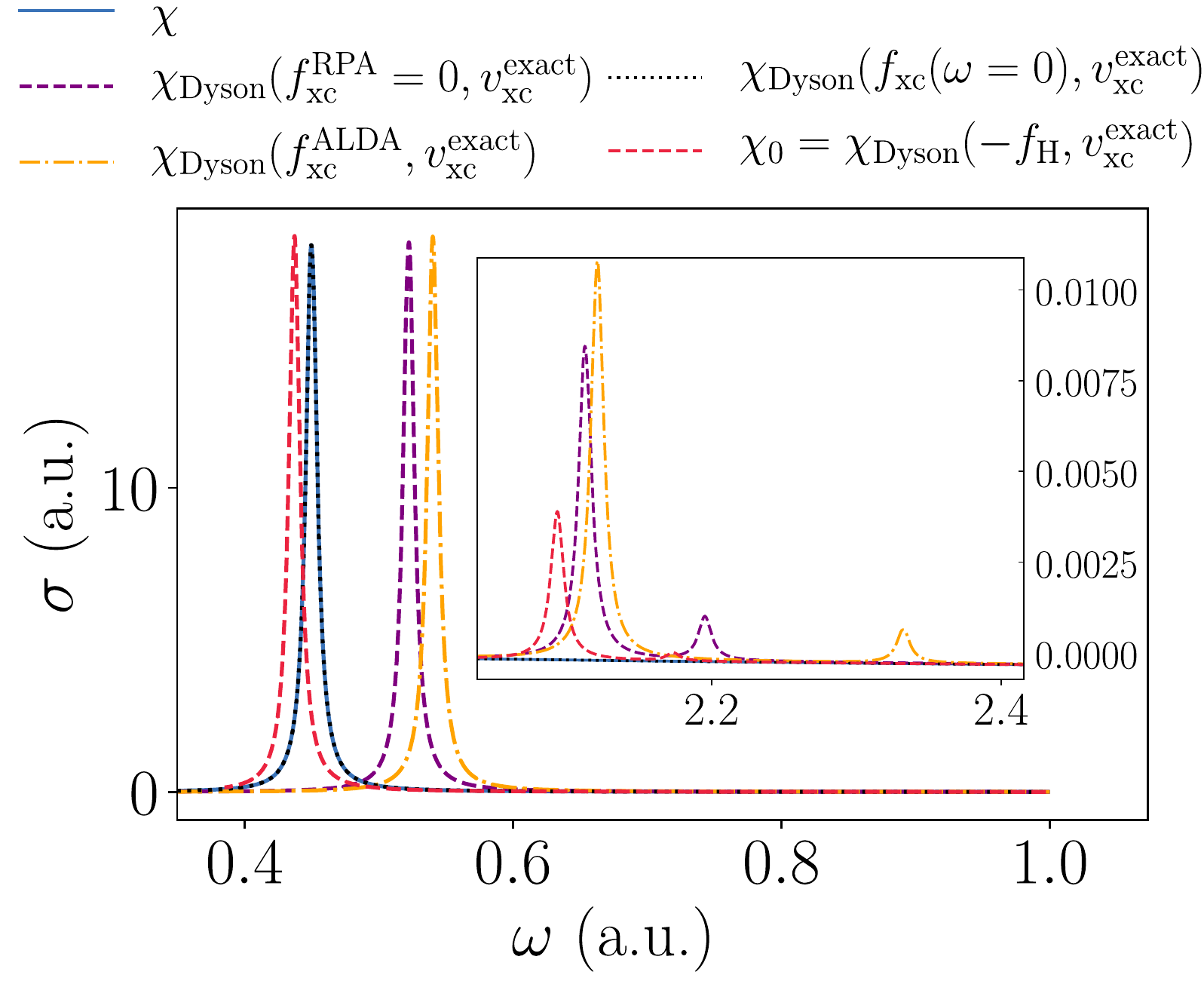}
\end{center}
\caption{The optical absorption spectrum for the quantum harmonic oscillator around the first transition, and around a chosen higher energy transition, calculated at various levels of approximation. The optical spectrum calculated from the interacting and Kohn-Sham response functions are plotted alongside the optical spectrum computed from the Dyson equation using the RPA, ALDA, and exact adiabatic xc kernels. At higher frequencies (inset), all optical peaks are suppressed for the quantum harmonic oscillator, a state of affairs that the exact adiabatic $f_\text{xc}$ reproduces, but the RPA and adiabatic LDA do not.}
\label{fig:opticalSpectrumQHO}
\end{figure}

To conclude matters for the quantum harmonic oscillator, we examine its optical absorption spectrum, and in particular highlight a failing of the $f_\text{xc}$ approximations considered in this work when compared to the exact and exact adiabatic xc kernels. The optical absorption spectrum, using the same range of approximations discussed in the case of the atom, is shown in Fig$.$ \ref{fig:opticalSpectrumQHO}. In the \textit{non-interacting} quantum harmonic oscillator, all transitions but the first are disallowed, otherwise known as its \textit{selection rules}. The inclusion of the Coulomb interaction lifts these special symmetries of the non-interacting quantum harmonic oscillator, but not enough for the previously disallowed transitions to be observed in the optical spectrum -- the dipole matrix elements for these \textit{allowed} transitions are $\mathcal{O}(10^{-8})$. On the other hand, the exact Kohn-Sham potential differs significantly from the harmonic form, which creates a series of visible peaks beyond the first in the optical spectrum; the transition rates for these transitions are vastly overestimated. 

In this situation, the RPA and adiabatic LDA do not achieve the required strong suppression of the optical peaks. Interestingly, the exact adiabatic kernel, as seen in Fig$.$ \ref{fig:QHOExactAdiabatic}, \textit{is} able to reproduce the exact optical spectrum across the entire frequency range considered $\omega \in [0,6]$ a.u., presumably due to the correct oscillator strengths used in its construction. This suggests that, in cases where the exact system possesses heavily suppressed transitions, perhaps due to symmetries that are not shared by the Kohn-Sham system, the typical $f_\text{xc}$ approximations are insufficient to recover the exact state of affairs, but improvement of their non-local spatial structure toward the exact adiabatic kernel can assist matters.

\subsection{Gauge freedom}
\label{sec:GaugeFreedom}

Two xc kernels that differ in structure that is predominantly captured with the gauge transform defined in Section \ref{sec:GaugeFreedomTheory} provides an explanation for the approximate agreement between the derived properties of the two xc kernels in question. This possibility is now considered, and in fact we shall demonstrate that the gauge freedom of $f_\text{xc}$ is \textit{not} sufficient to explain the similarity observed between, for example, the optical properties calculated using the adiabatic LDA and RPA in Section \ref{sec:Results}. Moreover, the particular form of the non-local spatial structure within the exact $f_\text{xc}(x,x',\omega)$ is in general not possible to capture with this gauge freedom, and it is unlikely that this line of reasoning is able to explain the efficacy of approximations of any kind.

In order to demonstrate this, an \textit{optimal gauge} is defined that transforms, in as much as is possible, one xc kernel into another using the gauge freedom, i.e. it brings one $f_\text{xc}$ toward another $f_\text{xc}$ in a particular matrix norm. The definition and derivation of the optimal gauge is given in the relevant section of the supplemental material. Furthermore, an illustration of the optimal gauge transform for each of the examples to follow is also provided in the supplemental material. The atom of Section \ref{sec:Results} is considered, and in particular the optimal gauge is found in order to match the RPA, $f_\text{xc}^\text{RPA} = 0$, with the adiabatic LDA used in this work, $f_\text{xc}^\text{ALDA}$ \citep{Entwistle2018}. As discussed in Section \ref{sec:GaugeFreedomTheory}, the vectors $g$ and $h$ introduce an unavoidable degree of non-locality, and the local spatial structure of the adiabatic LDA simply cannot be reproduced with a gauge transform of this kind applied to the RPA.

The story remains much the same when an attempt is made to match the adiabatic LDA xc kernel to the exact adiabatic xc kernel for the atom, $f_\text{xc}(x,x',\omega=0)$. The particular form of the spatial non-locality in the exact adiabatic xc kernel does not lend itself to the fairly inflexible structural freedom afforded by the functions $g$ and $h$. This conclusion holds true for \textit{almost} all the xc kernels examined in this work, namely, the spatial profile of the exact $f_\text{xc}$ at any $\omega$ is not possible to reproduce with a gauge transform applied to the adiabatic LDA or RPA, despite, at high frequencies, the similarity of the derived optical spectra from these approximations. 

A final remark is given in relation to the divergences in $f_\text{xc}$ studied in the previous section. The xc kernel around a divergence is approximately of the form of an outer product $|u\rangle \langle u|$, where $|u\rangle$ is the eigenvector of $f_\text{xc}$ that is diverging. The xc kernel is therefore mostly composed of $N$ degrees of freedom, a state of affairs that is \textit{a priori} much more amenable to the gauge freedom. In fact, the first pole in the $f_\text{xc}$ of the atom, i.e. the beginning of the non-adiabatic behavior, see panel (d) of Fig$.$ \ref{fig:pseudoatomicXCKernel}, is \textit{non-divergent} after the optimal gauge is applied. This divergence therefore \textit{does not} affect observable properties such as its associated optical peak, and indeed the exact adiabatic kernel is able to capture this peak, despite it existing squarely within the divergence. However, such a situation is not detected again for the remainder of the divergences. 

\subsection{Exchange-correlation potential versus exchange-correlation kernel}
\label{sec:GroundStateXC}

\begin{figure}[ht]
\begin{center}
\includegraphics[width=3.5in]{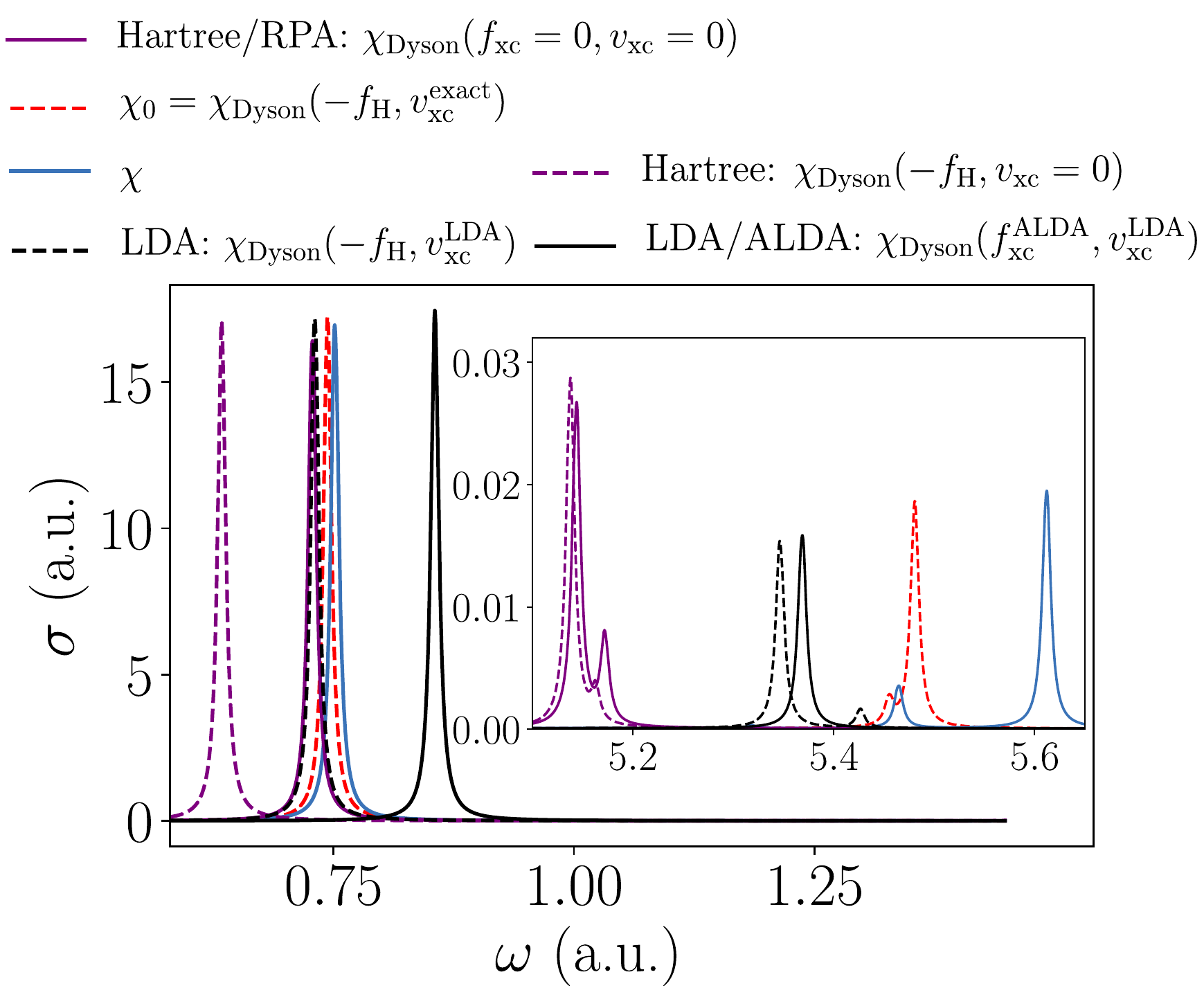}
\end{center}
\caption{The exact optical absorption spectra for the atomic system illustrated alongside the optical absorption spectrum from the non-interacting response functions $\chi_0$ calculated with wavefunctions and energies given by Hartree theory, an LDA, and the exact Kohn-Sham potential. The corrected transitions, given upon solution of the Dyson equation with the corresponding $f_\text{xc}$, are also shown for Hartree theory/RPA, and LDA/adiabatic LDA. The most accurate optical spectrum is that of the non-interacting exact Kohn-Sham system, and hence improving treatment of ground-state exchange-correlation toward this ideal is found to be most important. In particular, use of $f_\text{xc}$ to shift the non-interacting peaks toward the interacting peaks is, in general, unable to achieve accuracy improvements comparable to those observed from improving ground-state exchange-correlation -- this is seen most clearly at higher frequencies (inset).}
\label{fig:vxc_vs_fxc_OAS1}
\end{figure}

In finite systems, it is the conventional wisdom that an accurate ground-state xc potential is \textit{more} important for capturing the interacting excitation spectrum than a sophisticated $f_\text{xc}$ \citep{Ullrich2016,Botti2007,Marques2003,Marques2001,Petersilka2000}. The effect of an improved treatment of exchange and correlation in the ground state is shown in Fig$.$ \ref{fig:vxc_vs_fxc_OAS1}, which demonstrates the optical spectrum for the atomic system calculated from the \textit{non-interacting response function} $\chi_0$ at various levels of approximation. The transition energies and rates calculated from the exact Kohn-Sham system are considerably closer to the exact transitions than, for example, Hartree theory, which is increasingly poor at higher energies. 

These non-interacting response functions can be used, in conjunction with their \textit{corresponding} $f_\text{xc}$, to solve the Dyson equation, thus shifting the position and weight of the peaks. An xc kernel functional \textit{corresponds to} an xc potential functional if it is the second functional derivative of the xc potential with respect to the density. Thus, the RPA xc kernel corresponds to Hartree theory, and the adiabatic LDA xc kernel corresponds to the ground-state LDA from which it came. To use an xc kernel that does not correspond to the ground-state xc potential can violate various exact conditions \citep{Liebsch1985} -- this is evidently the case here for the zero-force sum rule. 

The aforementioned conventional wisdom is exhibited clearly in Fig$.$ \ref{fig:vxc_vs_fxc_OAS1}, namely, use of a corresponding $f_\text{xc}$ is only able to improve matters slightly beyond the non-interacting peaks, which is most visible at higher frequencies.  For this reason, even when using an incompatible $f_\text{xc}$, the calculation benefits from an improved treatment of ground-state exchange and correlation. This is evidenced in the case of the atom in Section \ref{sec:Results}, where using the RPA and adiabatic LDA xc kernels in conjunction with the exact Kohn-Sham ground-state yields a more accurate optical spectrum than if one were to attempt to keep the xc potential and xc kernel compatible. 

\section{Conclusions}

We have calculated and analyzed the spatial and frequency dependence of the exact xc kernel $f_\text{xc}$ of time-dependent DFT for three one-dimensional model systems: an atom, an infinite potential well, and a quantum harmonic oscillator. A set of numerical methods is designed to ensure numerical robustness. 

The xc kernel exhibits a significant non-local spatial structure at all frequencies, including at $\omega = 0$, i.e. the exact adiabatic $f_\text{xc}$. In lacking this structure, local approximations to $f_\text{xc}$ are found to be insufficient for recovering the lowest energy excitations, whereas the exact adiabatic xc kernel performs well. However, beyond the lowest few excitations, all the approximations considered here -- the exact adiabatic xc kernel, the RPA, and the adiabatic LDA -- are equally poor, and do not generally improve the optical spectrum obtained directly from the non-interacting Kohn-Sham response function $\chi_0$ (that is, setting $f_\text{xc} + f_\text{H}$ to zero everywhere). A notable exception is the quantum harmonic oscillator, whose optical transitions beyond the first are heavily suppressed, a feature that the exact adiabatic xc kernel is able to capture. In general, improvement of the spatial structure of adiabatic xc kernel approximations toward the exact adiabatic xc kernel is expected to assist matters for the lowest energy transitions, but beyond these transitions the lack of frequency dependence hinders all adiabatic xc kernels. In addition, the long-range limit of $f_\text{xc}$ for \textit{finite systems} $f_\text{xc} \rightarrow -f_\text{H}$ is confirmed, although the long-range character of $f_\text{xc}$ is demonstrated to be unimportant in the present context, in contrast to its character within and around the region of high density.

Drastic non-adiabatic behavior is observed in $f_\text{xc}$ for all systems studied in this work, and is, to a considerable extent, attributable to specific aspects of its analytic structure as a function of $\omega$. (Simple) poles in $f_\text{xc}$, related to certain interacting or non-interacting transitions that necessitate them, can in practice appear close to interacting excitations, for example, between two nearly degenerate charge-transfer excitations in a double-well system \cite{Note3}. It is possible that a gauge transform can remove the divergence in specific cases without affecting the optical spectrum, but this is the exception rather than the rule. If $f_\text{xc}$ is kept identical apart from removal of the diverging eigenvalue and its associated eigenvector, then $f_\text{xc}$ can be rendered unable to capture the relevant transition. This suggests that an $f_\text{xc}$ approximation that does not attempt to exhibit the non-adiabatic pole structure of the exact $f_\text{xc}$ cannot reproduce transitions with energies higher than the first few excitations. This is the case for single, double, triple, and so on, excitations alike. The fact that these divergences can be related to certain excitations that necessitate them provides a new perspective on the divergent character of $f_\text{xc}$ that is known to exist around double excitations \citep{Maitra2004,Gritsenko2009}. 

In general, the subtle spatial structure of the exact $f_\text{xc}$ cannot be captured by applying a gauge transformation to one of the usual kernel approximations. However, the divergent $\omega$-dependence discussed in the previous paragraph is more amenable to the gauge freedom. Indeed, in the case of the atom, the first divergence (around the third peak in the optical spectrum) turns out to be related to the exact adiabatic $f_\text{xc}$ with a gauge transform. Hence, the exact adiabatic approximation is able to describe the third peak in the optical spectrum, despite the significant frequency dependence of $f_\text{xc}$ around this peak. 

As noted earlier in this section, the simple non-interacting kernel $f_\text{xc} = - f_\text{H}$ often yields surprisingly good spectra, provided that the exact Kohn-Sham potential, or a good approximation to it, is used to calculate $\chi_0$. This is in part due to the fact that the exact Kohn-Sham transitions are, in certain circumstances, good approximations to the interacting transitions \citep{Petersilka2000}. By extension, in practical calculations, effort may be usefully devoted to improving approximations used for $f_\text{xc}$ and $v_\text{xc}$ individually, without an overriding need to maintain one as the functional derivative of the other. The quality of $v_\text{xc}$ is of particular importance, as previous authors have observed in specific cases \citep{Ullrich2016,Botti2007,Marques2003,Marques2001,Petersilka2000}.

In systems where predictive spectral accuracy beyond that provided by the simplest kernels is required, note should be taken of the intricate spatial non-locality and analytic structure as a function of $\omega$ exhibited by the exact kernels calculated in this paper. Approximate kernels that are spatially local (whether local-density or not), and/or exhibit no more than a gentle variation with $\omega$, are unlikely to prove adequate for calculating optical spectra and other aspects of the density response function. A fruitful direction appears to be kernels that are obtained by making a connection between the time-dependent DFT description and some level of many-body perturbation theory, such as the kernel obtained from the Bethe-Salpeter equation presented in \citep{Romaniello2009}, since non-locality and frequency dependence emerge automatically from even the simplest level of many-body perturbation theory.

\begin{acknowledgments}
The authors thank Michael Hutcheon and Matt Smith for helpful discussions. N.D.W is supported by the EPSRC Centre for Doctoral Training in Computational Methods for Materials Science for funding under grant number EP/L015552/1. 
\end{acknowledgments}

\bibliographystyle{apsrev4-2}
\bibliography{mainReferences}

\end{document}


\title{Insights from exact exchange-correlation kernels (Supplemental Material)}
\author{N. D. Woods}
\email{nw361@cam.ac.uk}
\affiliation{Theory of Condensed Matter, Cavendish Laboratory, University of Cambridge, Cambridge, CB3 0HE, United Kingdom}
\author{M. T. Entwistle}
\thanks{Present address: FU Berlin, Department of Mathematics and Computer Science, Arnimallee 6, 14195 Berlin, Germany}
\affiliation{Department of Physics, University of York, and European Theoretical Spectroscopy Facility, Heslington, York YO10 5DD, United Kingdom}
\author{R. W. Godby}
\affiliation{Department of Physics, University of York, and European Theoretical Spectroscopy Facility, Heslington, York YO10 5DD, United Kingdom}

\maketitle

\section{Methodology}
\subsection{Error and exact conditions}
\label{sec:ErrorAndExactConditions}

Having established two separate methods for regularizing the computation of the exact $f_\text{xc}$, we are able to discuss and quantify error in the present context. The foremost approach used to track error here is the mean absolute error between elements of the interacting response function $\chi$ and elements of the output of the Dyson equation $\chi_\text{Dyson}$ over the entire spatial and frequency grid. The quantity $\chi_\text{Dyson}$ is defined as
\begin{align}
\chi_\text{Dyson}(\omega) = \frac{\chi_0(\omega)}{I - \chi_0(\omega) (f_\text{xc}(\omega) + f_\text{H})}, \label{eq:finiteBasisDyson}
\end{align}
which, in exact arithmetic with an exact $f_\text{xc}$, is equal to the interacting response function. The \textit{method error} is thus understood as the difference between $\tilde{\chi}$ and $\chi$, and the \textit{numerical error} is the difference between $\tilde{\chi}$ and $\chi_\text{Dyson}$. The truncation parameters $b$ and $\bar{\lambda}$ defined in the main text are chosen such that the mean absolute error between $\chi$ and $\chi_\text{Dyson}$ is approximately equal to the method error, meaning little-to-no numerical error is present. Corroboration between both methods is also used as a means of extracting signal over noise.  \\

The xc kernel is known to be subject to a multitude of exact conditions \citep{Wagner2012}; these conditions have been important in both the historical development of $f_\text{xc}$ approximations \citep{Marques2006}, and also in other studies on the exact $f_\text{xc}$ \citep{Thiele2014,Thiele2009,Entwistle2019}. The xc kernels in this work are complex-symmetric and satisfy the Kramers-Kronig relations by construction. Exact conditions relating to the coupling constant dependence of $f_\text{xc}$ are not applicable here. The property that is most useful in the present context is the \textit{zero-force sum rule} \citep{Ullrich2012},
\begin{align}
\int_{-a}^a n(x') \frac{\partial}{\partial x'} f_\text{xc}(x,x',\omega) \ dx' = - \frac{\partial}{\partial x} v_\text{xc}(x), \label{eq:StrongZeroForce}
\end{align}
which states that perturbations in the xc potential cannot produce a net momentum over the entire domain at linear order, i.e. total momentum is conserved at zero. The extent to which the xc kernels computed here satisfy the zero-force sum rule can be seen in both a graphical and numerical form, which is shown in later sections.

\section{Results and discussion}
\subsection{Atom}

\begin{figure}[h!]
\begin{center}
\includegraphics[width=3in]{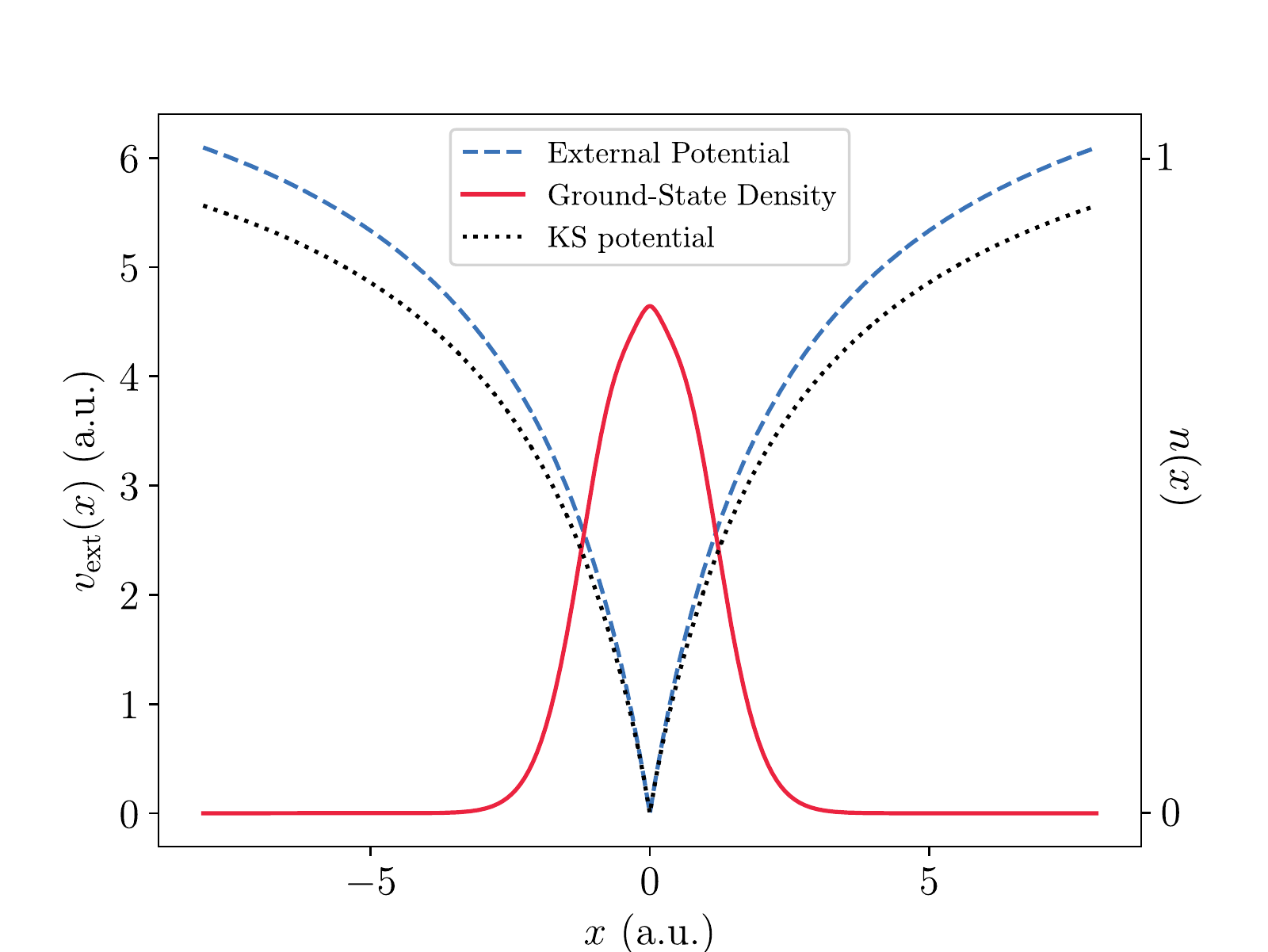}
\end{center}
\caption{The ground-state density, external potential, and reverse-engineered Kohn-Sham potential for the atomic system. The external and Kohn-Sham potential have been shifted for illustrative purposes.}
\label{fig:pseudoatom}
\end{figure}

\begin{figure}[h!]
\begin{center}
 \includegraphics[width=3.5in]{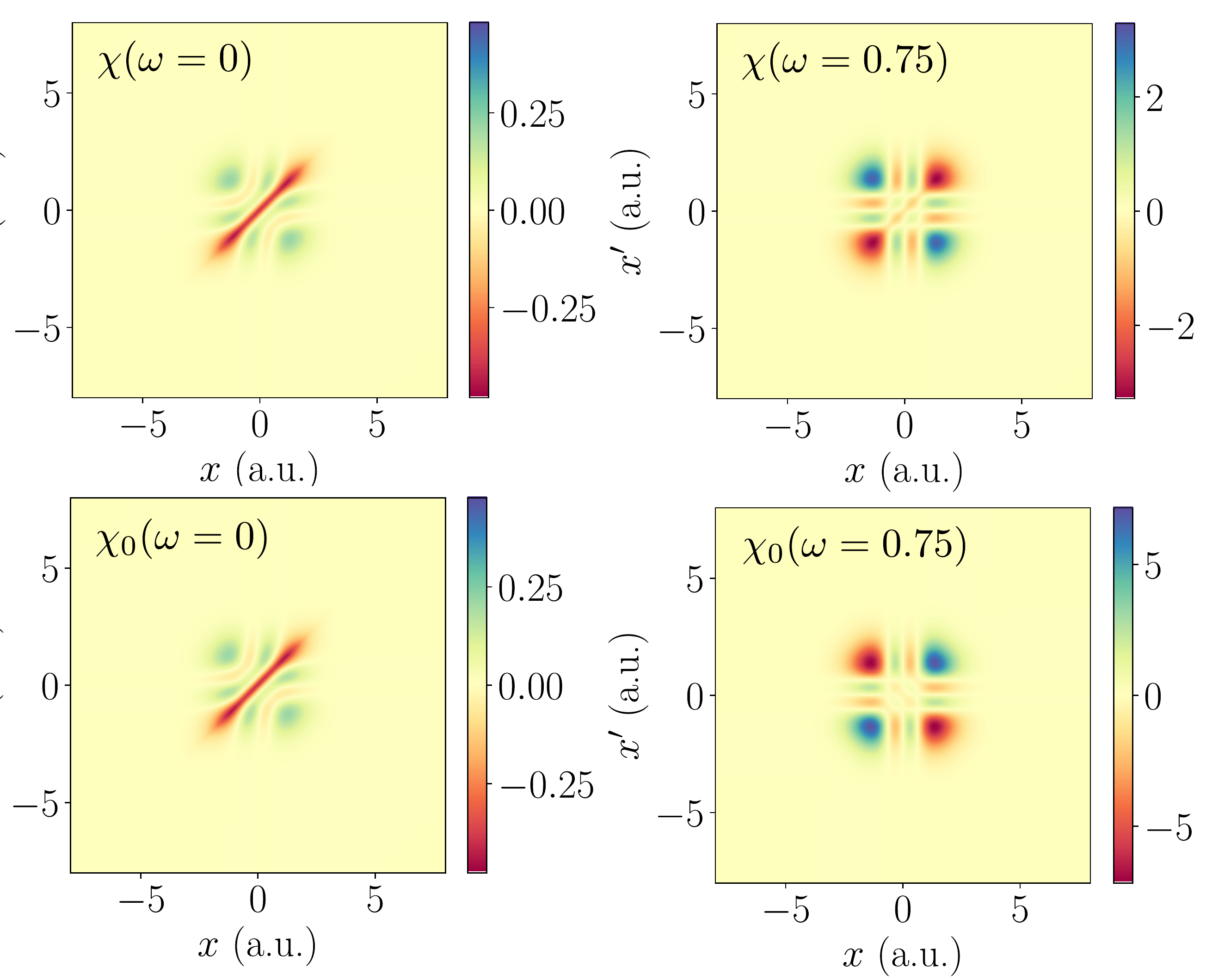}
\end{center}
\caption{The atomic response functions of the interacting $\chi$ (upper) and non-interacting $\chi_0$ (lower) systems at $\omega = 0$ a.u. and $\omega = 0.75$ a.u. (around the first interacting excitation).}
\label{fig:pseudoatomResponse}
\end{figure}

The atom is defined with the external potential
\begin{align}
v_\text{ext}(x) = \frac{-2}{|0.1x| + 0.5}.
\end{align}
The softening parameter $\alpha=1$ is used hereafter to regularize the Coulomb interaction (see main text). The electrons are confined to the region $x \in [-8,8]$ a.u., discretized over $N_x=151$ grid points. The interacting electron density, and its corresponding Kohn-Sham potential calculated with $\mathcal{O}(10^{-15})$ error, are given in Fig$.$ \ref{fig:pseudoatom}. The response functions are calculated using the Lehman representation for $\omega \in [0,6]$ a.u., discretized over $N_\omega = 2000$ grid points, see Fig$.$ \ref{fig:pseudoatomResponse}. In cases where the optical spectrum is required, a finite broadening $\eta = 0.005$ is used, meaning the response functions are non-Hermitian, and thus the real-space truncation method is used to calculate $f_\text{xc}$. \\

\subsubsection{Error}

The xc kernel is calculated using both the real-space truncation, and the eigenspace truncation, methods. Details of the real-space truncation are given in the main text -- the truncation parameter is set $b=5.8$ a.u., which is within a feasible range of $b$. The mean absolute error between the output of the Dyson equation, and the interacting response function, is
\begin{align}
||\chi - \chi_\text{Dyson}|| \coloneqq&  \sum_{(x,x',\omega) \text{ grid}} \frac{|\chi(x,x',\omega) - \chi_\text{Dyson}(x,x',\omega)|}{N_x^2 N_\omega} \nonumber \\
&\approx   10^{-9} \label{eq:pseudoatomMeanAbsError}.
\end{align}
As indicated in the background sections of the main text, the eigenspace truncation is more accurate than the real-space truncation, and is now used to demonstrate the difference between the method error and numerical error, which ultimately informs our choice of the truncation parameter(s). The method error, and numerical error, over a range of eigenspace truncation parameters $\bar{\lambda}$, are given in Table \ref{tbl:eigenspaceTruncationError}. The first row highlights the essence of the issue that this work aims to solve: setting the cutoff to machine precision, so that the effective null space no longer exists, gives us a `best possible' error of zero using exact arithmetic, but an excessively large error in finite precision -- an error $10^{-2}$ indicates that the reconstructed response function is poor beyond repair. Moving down the rows of Table \ref{tbl:eigenspaceTruncationError}, it can be seen that increasing the cutoff introduces only mild method error due to the effective null space of the response functions not overlapping. Furthermore, doing so has a significant regularizing effect whereby using the cutoff $\bar{\lambda} = 10^{-8}$ allowed the truncated response function to be reconstructed with a numerical error $\mathcal{O}(10^{-13})$, which is close to the method error. As expected, increasing the cutoff further begins to introduce significant method error that is matched well by the numerical error. The cutoff here is thus chosen to represent the best balance between these two errors, which is, in this instance, $\bar{\lambda} = 10^{-9}$ yielding an error $||\chi - \chi_\text{Dyson}||  \approx 10^{-11}$ (two orders of magnitude superior to the real-space truncation method). \\

\begin{table*}[t]\centering
\caption{The \textit{method error} $||\tilde{\chi} - \chi||$ is the difference between the truncated response function, and the interacting response function, and the \textit{numerical error} $||\tilde{\chi} - \chi_\text{Dyson} ||$ is the difference between the truncated response function, and the output of the Dyson equation. Both of these are shown here over a range of eigenspace truncation parameters $\bar{\lambda}$. Increasing the truncation parameter introduces error whose source is the approximation inherent to the truncation itself, but doing so has a significant regularizing effect, and the numerical error decreases accordingly. It is possible to choose a truncation parameter, e.g. $\bar{\lambda} = 10^{-8}$, such that both the method error \textit{and} numerical error are sufficiently low for the systems studied in this work.}
\begin{ruledtabular}
\begin{tabular}{ >{\centering\arraybackslash} m{4cm} >{\centering\arraybackslash} m{4.5cm} >{\centering\arraybackslash} m{5cm}} \label{tbl:eigenspaceTruncationError}
        Truncation Parameter $\bar{\lambda}$ &  Method Error $||\tilde{\chi} - \chi||$ &  Numerical Error $||\tilde{\chi} - \chi_\text{Dyson} ||$ \\[0.1cm]    \hline \\[-0.2cm]
       $1 \times 10^{-17}$ & $0.0000 \times 10^{0}$ & $3.4852 \times 10^{-2}$ \\
       $1 \times 10^{-14}$ & $7.7170 \times 10^{-18}$ &  $1.8161 \times 10^{-9}$ \\
       $1 \times 10^{-11}$ & $1.8947 \times 10^{-17}$ & $2.7201 \times 10^{-11}$ \\
       $1 \times 10^{-8}$ & $6.7188 \times 10^{-14}$ & $2.1082 \times 10^{-13}$ \\
       $1 \times  10^{-3}$ & $1.3512 \times 10^{-8}$ & $2.6115 \times 10^{-8}$ \\
\end{tabular}
\end{ruledtabular}
\end{table*}

\begin{figure}[h!]
\begin{center}
\includegraphics[width=3.5in]{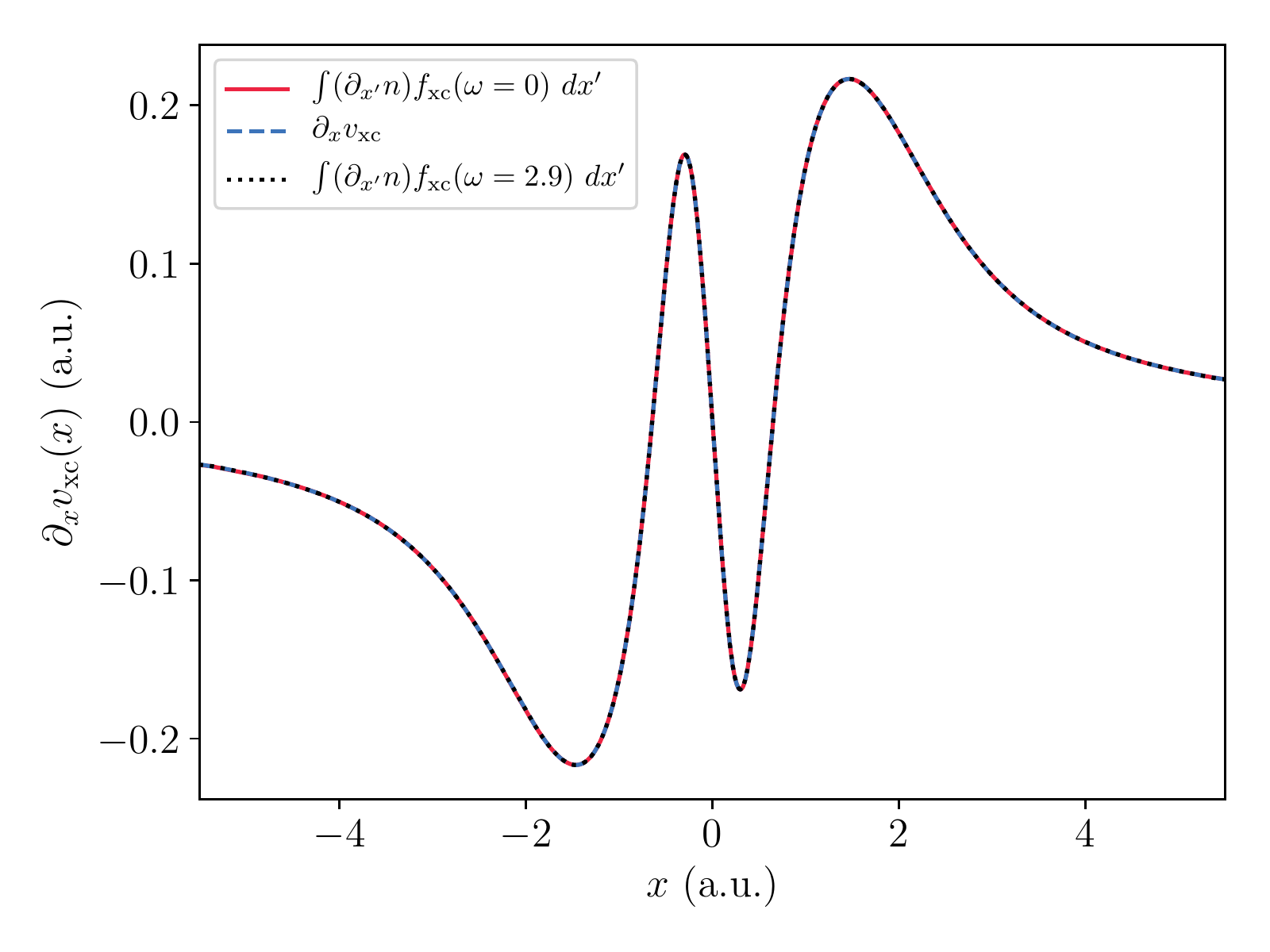}
\end{center}
\caption{Agreement between the left-hand and right-hand sides of the zero-force sum rule, Eq$.$ (\ref{eq:StrongZeroForce}), for the exact atomic numerical $f_\text{xc}$ at $\omega = 0$ a.u. and $\omega = 2.9$ a.u.}
\label{fig:pseudoatomZFSR}
\end{figure}

There are various avenues available to demonstrate the accuracy of the numerical xc kernels, some of which are detailed in Section \ref{sec:ErrorAndExactConditions}. First, the zero-force sum rule is calculated across the the entire $\omega$-range, and gives an average error between the left- and right-hand side in Eq$.$ (\ref{eq:StrongZeroForce}) of $\mathcal{O}(10^{-6})$ \textit{within the inner region determined by $b$}, this agreement is illustrated in Fig$.$ \ref{fig:pseudoatomZFSR}. Outside of this region, which is not shown, the error in the zero-force condition is significant, but the outer region contributes little to recovering the exact response function and other derived properties. Since $f_\text{xc}$ is not expressed as a closed-form functional of the density, we are prevented from studying the coupling constant dependence of $f_\text{xc}$, and thus also from examining the exact conditions on $f_\text{xc}$ relating to the adiabatic connection -- doing so would require a series of cumbersome numerical calculations, subject to errors of their own. The optical spectra given by $\chi$ and $\chi_\text{Dyson}$, as shown in the main text, also match across the energy range studied, which further evidences the fact that the truncation did not change observable properties of interest, while maintaining numerical stability. \\

\subsubsection{Anchoring the exchange-correlation kernels}

\begin{figure}[h!]
\begin{center}
\includegraphics[width=3in]{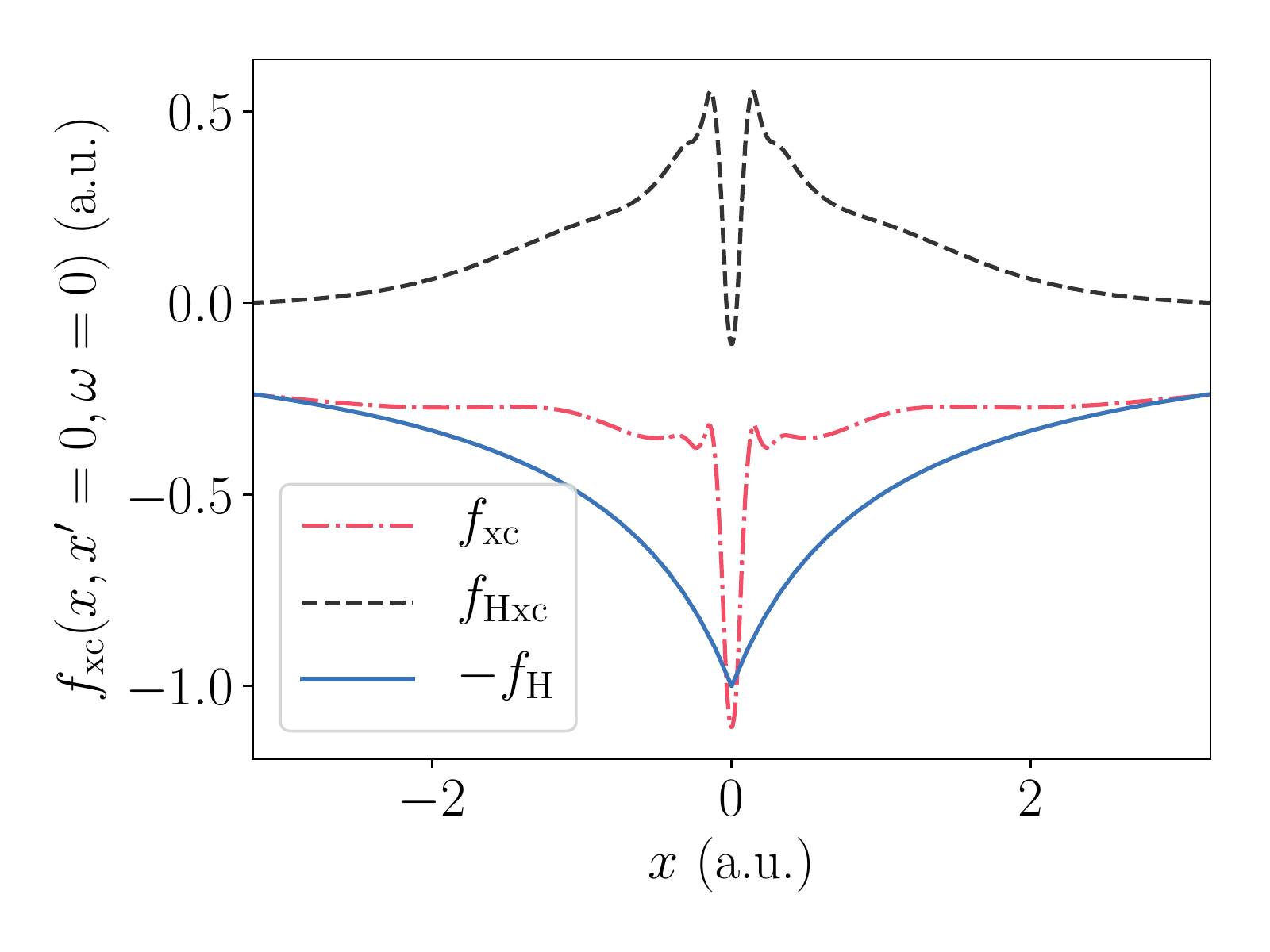}
\includegraphics[width=3in]{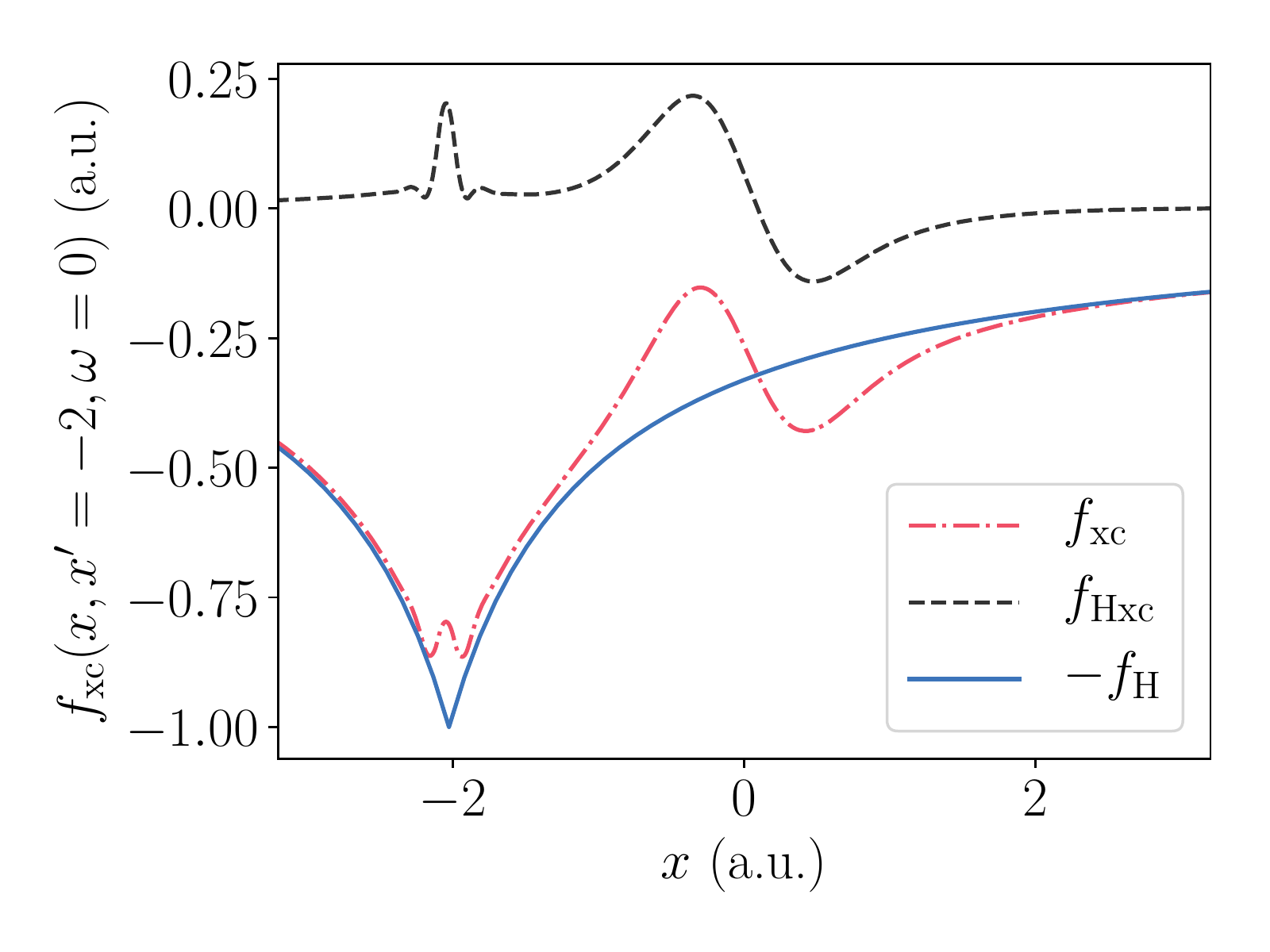}
\end{center}
\caption{The exact adiabatic xc kernel $f_\text{xc}(x,x',\omega=0)$ for the atom along $x'=0$ a.u. (top) and $x'=-2$ a.u. (bottom). The xc kernel (red dot-dash), the Hxc kernel (black dash), and the Hartree kernel (blue solid) are shown, and the long-range limit $f_\text{xc} \rightarrow -f_\text{H}$ is observed.}
\label{fig:pseudoatomLongRange}
\end{figure}

As described in the main text, the appropriate additive constant shift of $f_\text{xc}$ is determined by imposing the long-range limit $f_\text{xc} + f_\text{H} \rightarrow 0$. Fig$.$ \ref{fig:pseudoatomLongRange} shows that this limit is achieved reliably for the atom.

\subsubsection{Optical spectrum and transition energies}

In the main text, the optical spectrum is used to evaluate both the exact $f_\text{xc}$, and a variety of $f_\text{xc}$ approximations. In particular, we elaborate here on the methods used to establish a correspondence between a non-interacting transition, and an interacting transition. In Table \ref{tbl:pseudoatomStateOverlap}, the overlap of the final states involved in an exact transition, and its corresponding Kohn-Sham transition, is given. It can be seen that the interacting system is of sufficient single-particle character to make an identification between transitions. Not all excitations are visible excitations, i.e. excitations that show up in the optical spectrum, and this is true of the first double excitation, that appears in the interacting system $|\Psi_0\rangle \rightarrow |\Psi_{14}\rangle$ with energy $\omega = 3.7065$ a.u.

\begin{table*}[t]\centering
\caption{Identification of interacting and Kohn-Sham transitions for the atom using the overlap of the interacting and Kohn-Sham final states as a measure. This includes the first double excitation at $\omega = 3.5594$ a.u. that is not visible in the optical spectrum.}
\begin{ruledtabular}
\begin{tabular}{ >{\centering\arraybackslash} m{4cm} >{\centering\arraybackslash} m{4.5cm} >{\centering\arraybackslash} m{3cm} >{\centering\arraybackslash} m{3cm}  >{\centering\arraybackslash} m{3cm}} \label{tbl:pseudoatomStateOverlap}
       Exact Transition &  Kohn-Sham transition & Appears in Optical Spectrum &  Overlap in Final State & Exact vs. KS transition energy (a.u.) \\[0.1cm]
       \hline \\[-0.2cm]
      $| \Psi_0\rangle \rightarrow |\Psi_1\rangle$ & $\vert  \Phi_{(0,1)}\rangle \rightarrow | \Phi_{(0,2)}\rangle$ & \cmark & 0.9996 & 0.7431 \ \ \ 0.7513 \\
      $|\Psi_0\rangle \rightarrow |\Psi_2\rangle$ & $\vert  \Phi_{(0,1)}\rangle \rightarrow | \Phi_{(0,3)}\rangle$ & \xmark & 0.9993 & 1.3230 \ \ \ 1.3291  \\
      $|\Psi_0\rangle \rightarrow |\Psi_3\rangle$ & $\vert  \Phi_{(0,1)}\rangle \rightarrow | \Phi_{(0,4)}\rangle$ & \cmark & 0.9995 & 1.7274 \ \ \ 1.7325  \\
       $| \Psi_0\rangle \rightarrow |\Psi_4\rangle$ & $\vert  \Phi_{(0,1)}\rangle \rightarrow | \Phi_{(0,5)}\rangle$ & \xmark & 0.9987 & 2.0721 \ \ \ 2.0762 \\
       $|\Psi_0\rangle \rightarrow |\Psi_5\rangle$ & $\vert  \Phi_{(0,1)}\rangle \rightarrow | \Phi_{(1,2)}\rangle$ & \cmark & 0.9978 & 2.2384 \ \ \ 2.3118 \\
       $| \Psi_0\rangle \rightarrow |\Psi_9\rangle$ & $\vert  \Phi_{(0,1)}\rangle \rightarrow | \Phi_{(1,3)}\rangle$ & \cmark & 0.9842 & 2.8183 \ \ \ 2.9398 \\
       $| \Psi_0\rangle \rightarrow |\Psi_{14}\rangle$ & $\vert  \Phi_{(0,1)}\rangle \rightarrow | \Phi_{(2,3)}\rangle$ & \xmark & 0.9917 & 3.5594 \ \ \ 3.7065 \\   
       $| \Psi_0\rangle \rightarrow |\Psi_{18}\rangle$ & $\vert  \Phi_{(0,1)}\rangle \rightarrow | \Phi_{(1,7)}\rangle$ & \cmark & 0.9791 & 4.2339 \ \ \ 4.1056 \\    
       $| \Psi_0\rangle \rightarrow |\Psi_{19}\rangle$ & $\vert  \Phi_{(0,1)}\rangle \rightarrow | \Phi_{(0,12)}\rangle$ & \cmark & 0.9814 & 4.4382 \ \ \ 4.4301 \\ 
\end{tabular}
\end{ruledtabular}
\end{table*}

\subsection{Infinite potential well}

\begin{figure}[h!]
\begin{center}
\includegraphics[width=3in]{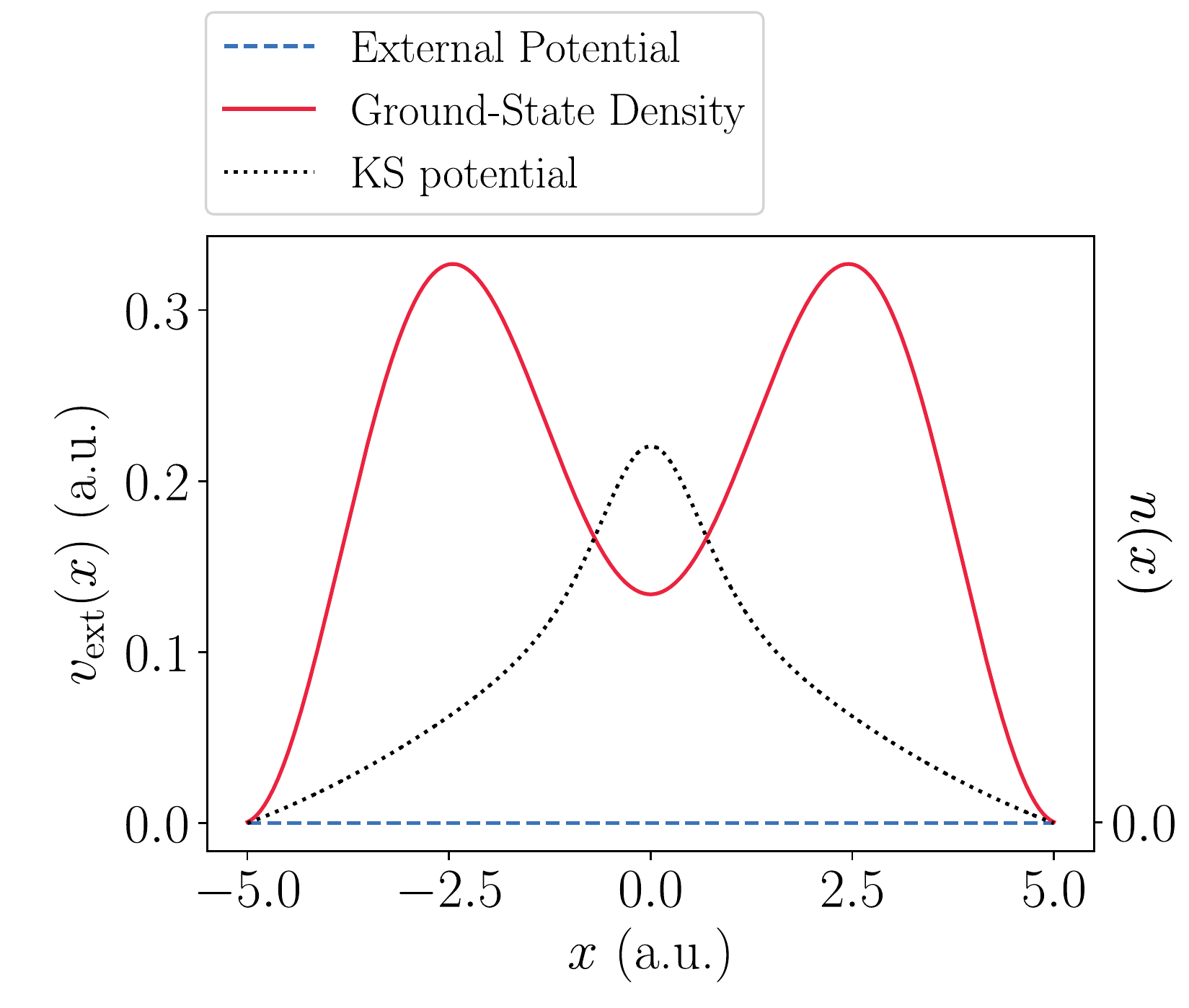}
\end{center}
\caption{The ground-state density, external potential, and reverse-engineered Kohn-Sham potential for the infinite potential well. The external and Kohn-Sham potential have been shifted for illustrative purposes.}
\label{fig:infPotWell}
\end{figure}

The infinite potential well is defined with the external potential
\begin{align}
v_\text{ext}(x) = 0.
\end{align}
The electrons are confined to the region $x \in [-5,5]$ a.u., discretized over $N_x=121$ grid points. The interacting electron density, and its corresponding Kohn-Sham potential calculated with $\mathcal{O}(10^{-15})$ error, are given in Fig$.$ \ref{fig:infPotWell}. The response functions are calculated using the Lehman representation for $\omega \in [0,6]$ a.u., discretized over $N_\omega = 2000$ grid points. The infinite potential well, as Fig$.$ \ref{fig:infPotWell} illustrates, has no region of near-zero density toward the outer parts of the domain, and therefore the response functions do not succumb to numerical ill-conditioning apart from due to the constant shift $c$ which has eigenvalue zero. This eigenvalue and its eigenvector are discarded within the singular value decomposition (SVD). Excitations of the infinite potential well are valid up to arbitrary energy, since the external potential coincides with the boundary conditions of the simulation. 

\subsubsection{Divergences and zeros of the response functions}

\begin{figure}[h!]
\begin{center}
\includegraphics[width=3in]{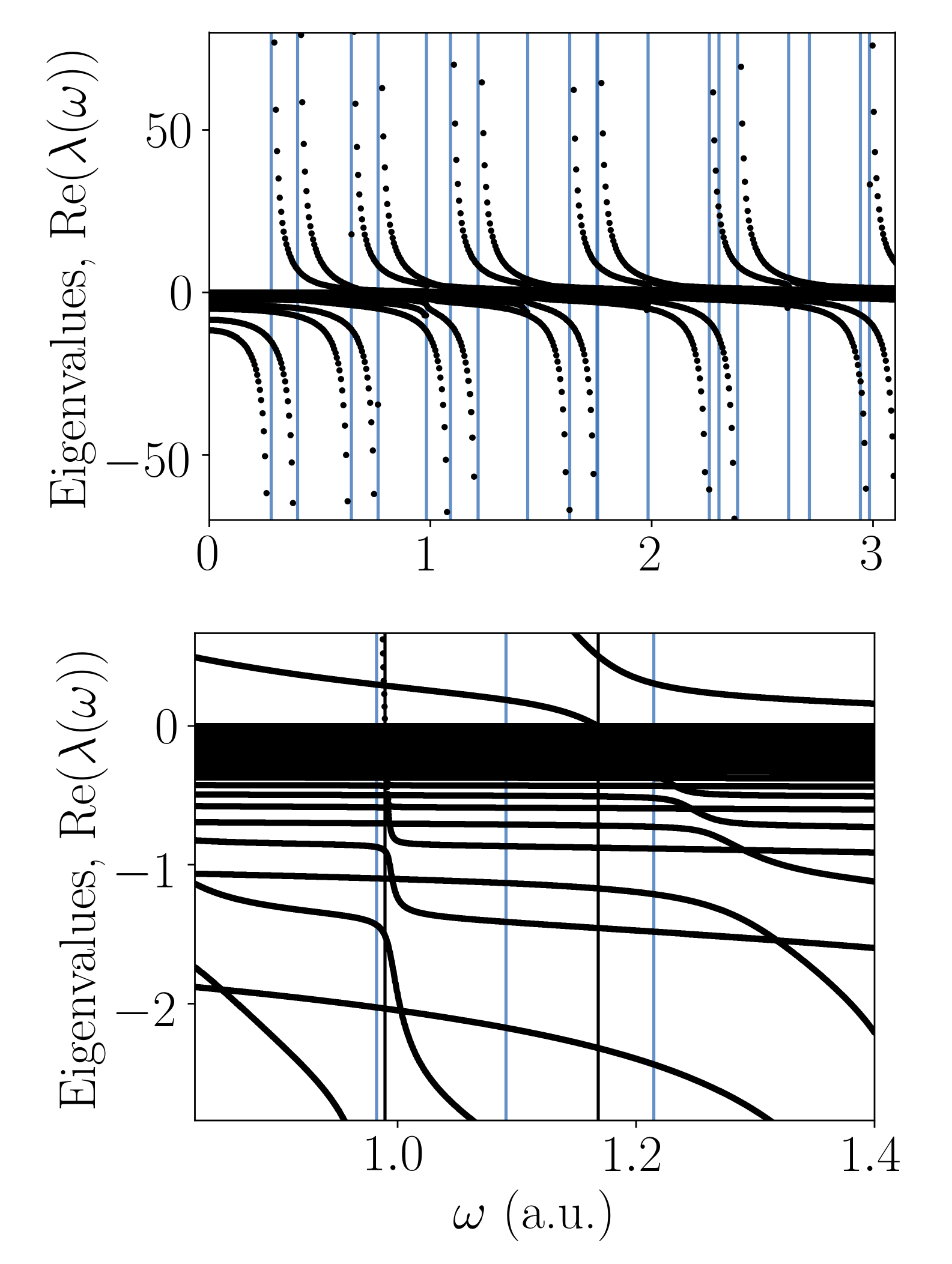}
\end{center}
\caption{Eigenvalues of the infinite potential well interacting response function $\chi$ (black dots), which are shown to diverge around excitation energies (blue lines). The lower panel zooms in on the range $\omega \in [0.8,1.4]$ a.u., where the response function crosses zero twice (black lines), which manifest as singularities in $f_\text{xc}$.}
\label{fig:infPotWellChiEigenvalues}
\end{figure}

As established in the main text, the predominant source of non-adiabatic behavior in $f_\text{xc}$ is due to zero eigenvalues of the response functions. It is possible here to elaborate on this fact, and introduce some of the analysis tools used in the main text. Let us first consider a plot of the eigenvalues of the interacting response function $\chi$, see Fig$.$ \ref{fig:infPotWellChiEigenvalues}. As expected, there are singularities, \textit{simple poles}, in the interacting response function at frequencies that correspond to interacting transition energies. The response function in the vicinity of the $n^\text{th}$ excitation behaves as
\begin{align}
\chi(x,x',\omega) &\approx  \frac{\langle \Psi_0 | \hat{n}(x) | \Psi_n \rangle \langle \Psi_n | \hat{n}(x') | \Psi_0  \rangle}{\omega - \Omega_n} \\
& \coloneqq \frac{f_n(x) f_n(x')}{\omega - \Omega_n},
\end{align}
where $f_n(x)$ is given the name \textit{excitation function}. In a finite basis, these excitation functions are $N$-dimensional vectors, and the response function around the divergence has eigenvector $|f_n\rangle$ with eigenvalue $(\omega - \Omega_n)^{-1}$, i.e.
\begin{align}
\chi(\omega) \approx \frac{|f_n\rangle \langle f_n|}{\omega - \Omega_n}. \label{eq:ChiAroundEx}
\end{align}
As shown in the main text, and in the lower panel of Fig$.$ \ref{fig:infPotWellChiEigenvalues}, zooming between the excitations reveals that the eigenvalues of the response matrices as a function of frequency, $\lambda_i(\omega)$, cross zero, which is established for non-interacting response functions in \citep{Mearns1987}, but shown here for \textit{interacting} response functions for the first time. At such frequencies, the response function is non-invertible, and $f_\text{xc}$ does not exist, i.e. $f_\text{xc}$ has an infinitely difficult job reconstructing $\chi$ from $\chi_0$ in the direction corresponding to the eigenvalue that is zero, which can be gathered from inspection of the Dyson equation.

\begin{figure}[h!]
\begin{center}
\includegraphics[width=3in]{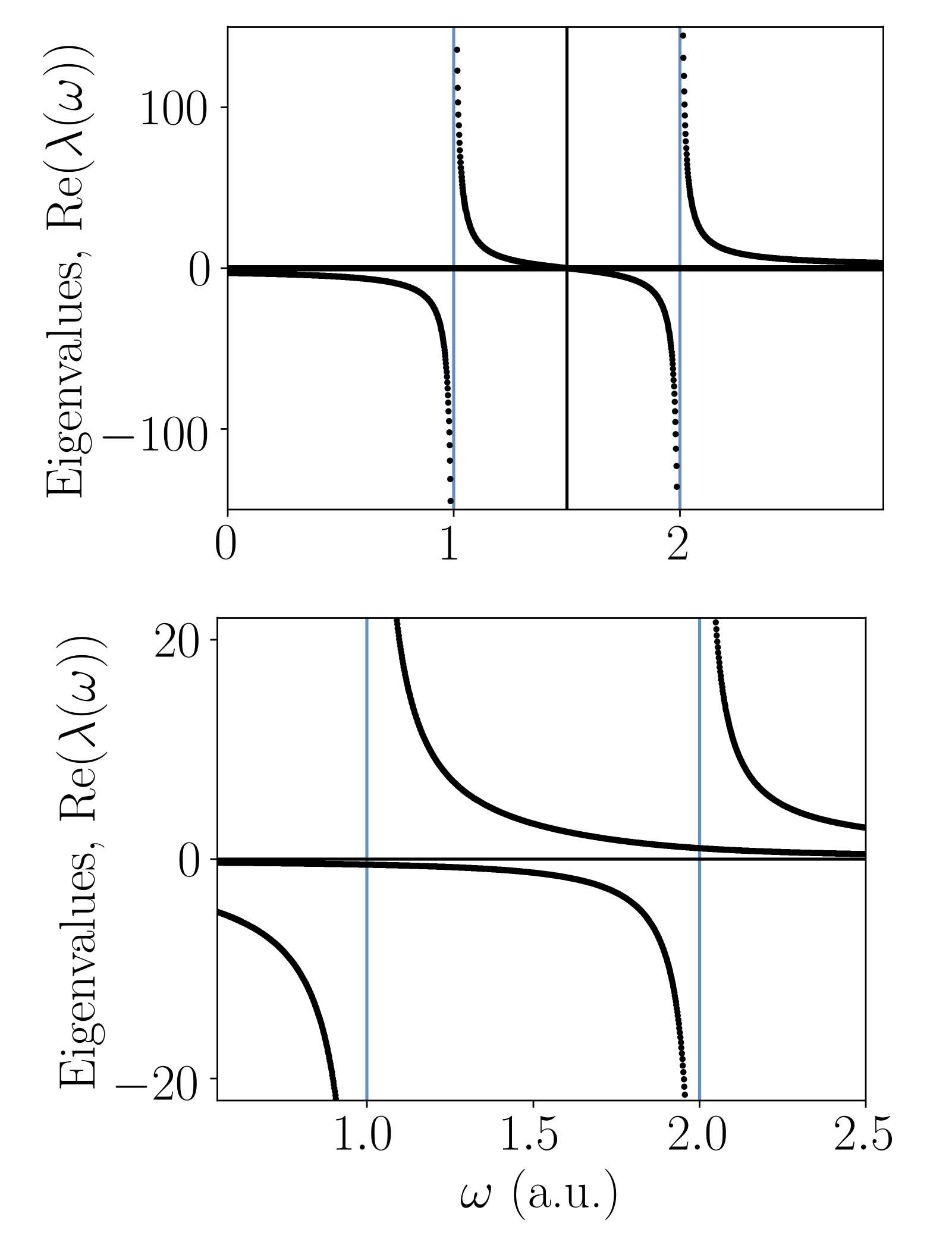}
\end{center}
\caption{Eigenvalues of the two-state model $\chi$ with excitation energies $\Omega_1 = 1$ a.u. and $\Omega_2 = 2$ a.u. In the upper panel, the excitation functions, $|f_1\rangle$ and $|f_2\rangle$, are parallel, in which case, a zero (black line) exists at the geometric mean of the excitation energies (blue lines), $\omega = \sqrt{2}$ a.u. In the lower panel, the excitation functions are given overlap $\alpha = 0.5$, which results in an avoided crossing about the $\omega$-axis, and thus an avoided zero.}
\label{fig:modelChiEigenvaluesV1}
\end{figure}

A \textbf{two-state model response function} can provide insight into the source of these zeros, and more generally elucidate the features observed in the lower panel of Fig$.$ \ref{fig:infPotWellChiEigenvalues}, see \citep{Entwistle2018} for a scalar version of the model to follow. The two-state model is established by supposing that there are two excitations with energies $\Omega_n$ and $\Omega_m$ whose excitation functions are orthogonal to the rest, $\langle f_n |f_i\rangle = 0$ and $\langle f_m |f_i\rangle = 0$ for $i \neq n, m$. The (normalized) excitation functions are given some finite overlap
\begin{align}
\langle f_n | f_m \rangle = \alpha,
\end{align}
parameterized by $\alpha \in [0,1]$. Each excitation contributes a term to the response function in the form of Eq$.$ (\ref{eq:ChiAroundEx}), and its $\omega \rightarrow -\omega$ counterpart. The eigenvalues of $\chi$ relating to the $2 \times 2$ subspace that has been isolated are
\begin{align}
\lambda_{\pm}(\omega) =& x(\omega) + y(\omega) (1 - 2 \alpha + 2 \alpha^2) \pm  \label{eq:TwoStateEigenvalues} \\
&\sqrt{(-x - y (1 - 2 \alpha + 2 \alpha^2)^2 - 4 xy (1-2\alpha + \alpha^2)},  \nonumber 
\end{align}
where
\begin{align}
x(\omega) & \coloneqq  \frac{1}{\omega - \Omega_n} - \frac{1}{\omega + \Omega_n} \\
y(\omega) & \coloneqq  \frac{1}{\omega - \Omega_m} - \frac{1}{\omega + \Omega_m}.
\end{align}
Setting Eq$.$ (\ref{eq:TwoStateEigenvalues}) equal to zero reveals that this two-state model crosses zero only when the response functions are parallel, i.e. $\alpha = 1$, and the frequency at which the zero occurs is equal to the geometric mean of the two excitation energies, $\omega = \sqrt{\Omega_n \Omega_m}$, see upper panel of Fig$.$ \ref{fig:modelChiEigenvaluesV1}.

If the excitation functions have some non-integer overlap $\alpha \neq 1$, which is generally the case in practice, the eigenvalues exhibit an \textit{avoided crossing} about the $\omega$-axis, and thus never cross zero, see lower panel of Fig$.$ \ref{fig:modelChiEigenvaluesV1}. Such behavior is displayed clearly in the eigenvalues of the exact response function, Fig$.$ \ref{fig:infPotWellChiEigenvalues}, however, the avoided crossings appear below the $\omega$-axis. This is accounted for in the two-state model by supposing there exist higher energy excitations $\Omega_i >> \Omega_{n/m}$ with excitation functions $|f_i\rangle$ that overlap with the excitation functions of interest, $|f_n\rangle$ and $|f_m\rangle$. Since these excitations are much higher in energy, their contribution to the response function $\chi$ \textit{around the n$^\text{th}$ and m$^\text{th}$ excitation} is given by subtracting a constant $c$ to all components of the $2 \times 2$ subspace (derivation omitted). 

\begin{figure}[h!]
\begin{center}
\includegraphics[width=3in]{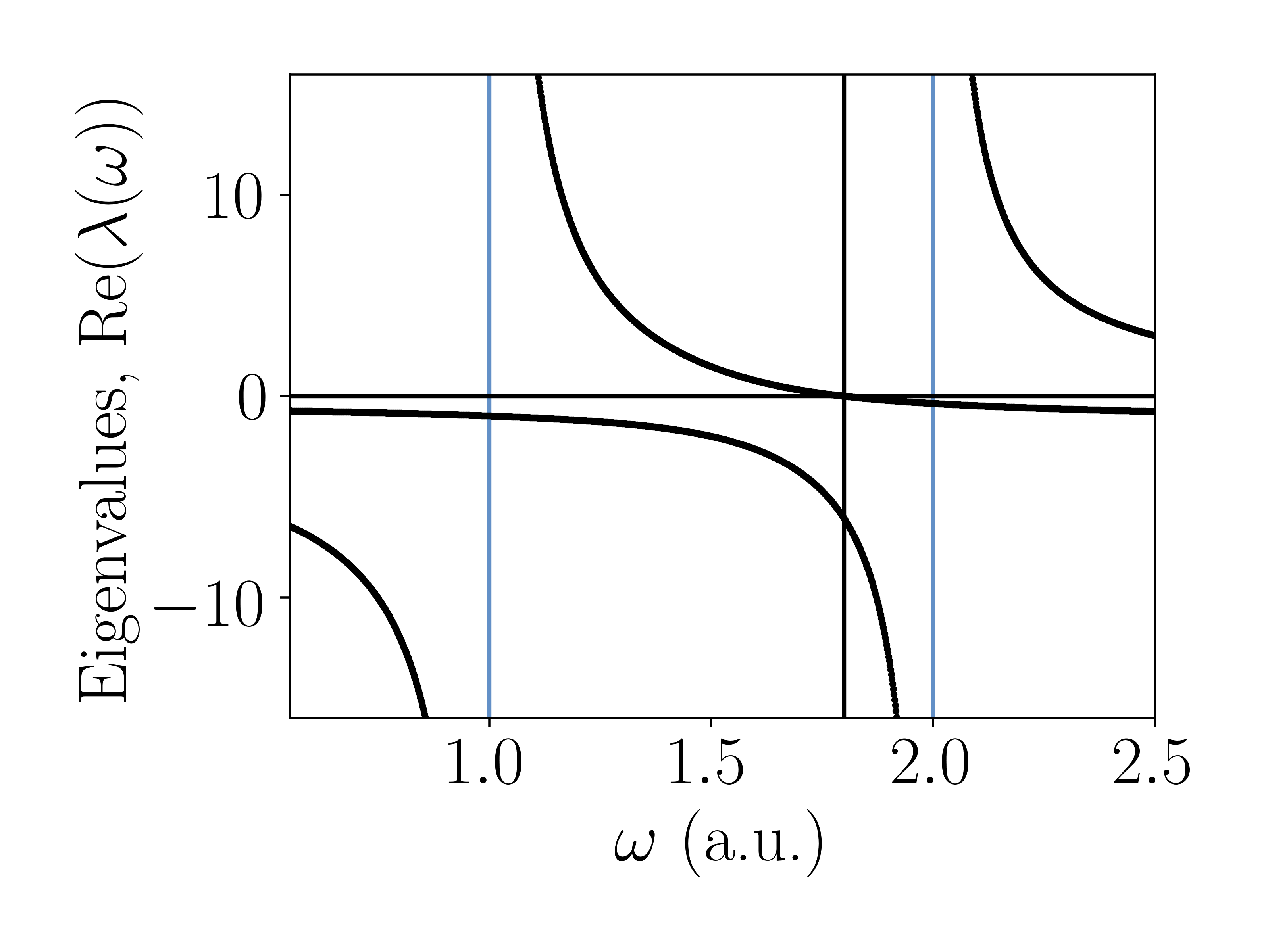}
\end{center}
\caption{Eigenvalues of the two-state model $\chi$ from Fig$.$ \ref{fig:modelChiEigenvaluesV1} with additional excitations at much higher energy, parameterized with $c=0.9$ and $\alpha = 0.8$ (see main text). The excitations of higher energy serve to shift the eigenvalues down, allowing a zero (black line) in $\chi$ (singularity in $f_\text{xc}$) to occur where it was otherwise prohibited in the lower panel of Fig$.$ \ref{fig:modelChiEigenvaluesV1}.}
\label{fig:modelChiEigenvaluesV2}
\end{figure}

As Fig$.$ \ref{fig:modelChiEigenvaluesV2} demonstrates, this modification of the two-state model, in part, leads to a negative shift in the eigenvalues of Eq$.$ (\ref{eq:TwoStateEigenvalues}), which in turn allows $\chi$ to cross zero and exhibit an avoided crossing \textit{below} the $\omega$-axis -- this is the behavior observed in the eigenvalues of the exact response function. Therefore, zeros in either the exact or Kohn-Sham response function occur between two isolated excitations whose excitation functions overlap, and this is the sense in which a zero can be related to certain excitations. The distinction made in the main text between \textit{paired} and \textit{unpaired} singularities in $f_\text{xc}$, i.e. singularities relating to single- and multi-excitations respectively, should now come as no surprise.

Note that the zeros in the interacting and Kohn-Sham response functions appear at different frequencies, and thus, at these frequencies, the difference of the inverses $\chi^{-1} - \chi_0^{-1}$ diverges, and so does $f_\text{xc}$. The importance of these divergences is not obvious to determine on analytic grounds; the divergences are a significant departure from the adiabatic limit, but are they required to recover optical properties?

It is argued, for example, in \citep{VanLeeuwen2001}, that the singularity itself appears between excitations, and thus is unlikely to be of concern, except in the case where a singularity is condensed between two nearly degenerate excitations. This is contested here on three grounds: first, nothing prevents a zero in $\chi_0$ from occurring at an excitation in $\chi$ (and vice-versa), and while such an occurrence might seem improbable, it almost happens without design in the infinite potential well at $\omega = 1.025$ a.u. (see the Fig$.$ 7 in the main text). Second, the behavior \textit{around} the singularity in $f_\text{xc}$, whose source is the singularity, can be (and is) large in magnitude and of importance, even if the singularity itself is not. Third, in a finite basis, it is possible in principle that a zero of $\chi$ can exist on a pole (excitation) in $\chi$. This final example can be realized within the $\alpha = 1$ two-state model by including an additional excitation with two properties: 1) its energy is equal to the geometric mean of the original excitation energies, e.g. those in Fig$.$ \ref{fig:modelChiEigenvaluesV1} (upper panel), 2) its excitation function is orthogonal to the original two \footnote{The orthogonality of the excitation function implies that the singularity in $f_\text{xc}$ at the excitation is not relevant for capturing the excitation. However, this example serves to demonstrate a more general feature, the rarity of which in practice is unknown.}. 

\subsubsection{Removal of divergences in the exchange-correlation kernel}

The central question is the following: \textbf{are these divergences in $f_\text{xc}$ required in order to capture interacting excitations?} The method used here to answer this question is to consider an $f_\text{xc}$ that is identical to the exact $f_\text{xc}$ except with the diverging eigenvalue and eigenvector removed using an SVD \footnote{In the case of a non-Hermitian $f_\text{xc}$, the solution is reformulated with projection operators, rather than an SVD.}. In other words, the divergent $f_\text{xc}$ around a singularity has the eigendecomposition
\begin{align}
f_\text{xc}(\omega)|u_i (\omega)\rangle = \lambda_i(\omega) |u_i (\omega)\rangle,
\end{align}
where the response functions, and hence also $f_\text{xc}$, are Hermitian ($\eta = 0$).

\begin{figure}[h!]
\begin{center}
\includegraphics[width=3in]{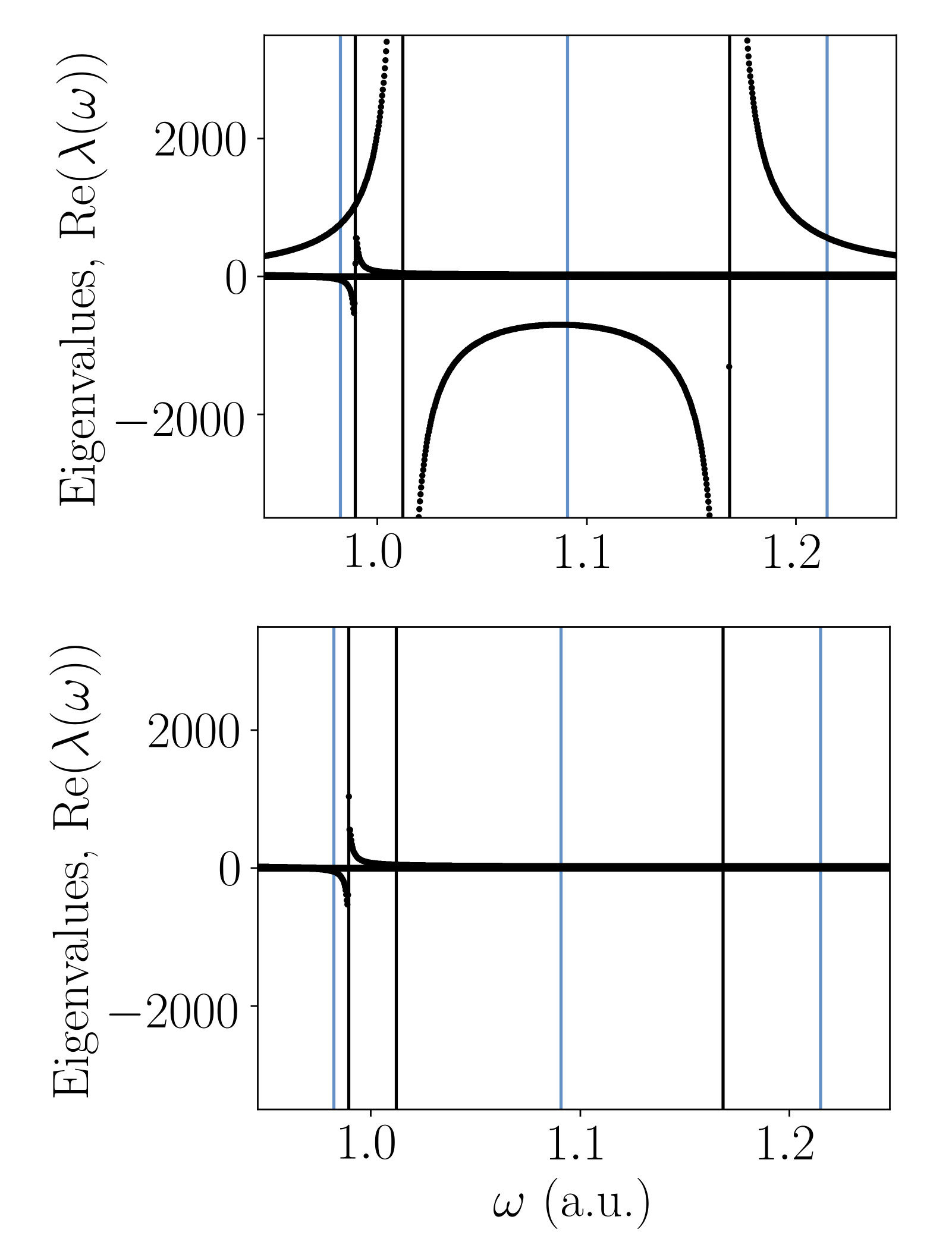}
\end{center}
\caption{Eigenvalues of $f_\text{xc}$ for the infinite potential well as a function of frequency, labeled Re($\lambda(\omega)$) (upper panel). Within the frequency range shown, the predominant non-adiabatic behavior of $f_\text{xc}$ is a result of three singularities (black lines) at $\omega = 0.99, 1.01, 1.17$ a.u. The lower panel demonstrates the $f_\text{xc}$ that has identical eigendecomposition apart from the divergences related to the singularities at $\omega = 1.01, 1.17$ a.u. in the upper panel.}
\label{fig:infPotWellfxcEigenvalues}
\end{figure}

At a divergence in $f_\text{xc}$, there is an eigenvalue that is larger in magnitude than the rest, let us define this as eigenvalue as $|\lambda_0| >> |\lambda_{i \neq 0}|$. The relevant figure from the main text is reproduced here for illustrative purpose, Fig$.$ \ref{fig:infPotWellfxcEigenvalues} (upper panel), where the eigenvalue $\lambda_0(\omega)$ in the range $\omega \in [1,1.2]$ a.u. is apparent. To remove this eigenvalue, and its corresponding eigenvector, $|u_0(\omega)\rangle$, from $f_\text{xc}$ gives it the eigenvalues depicted in Fig$.$ \ref{fig:infPotWellfxcEigenvalues} (lower panel). This makes it clear that, even though the interacting excitation at $\omega = 1.08$ a.u. is not \textit{on} a singularity, the divergent eigenvalue, i.e. the eigenvalue that is larger than the rest in magnitude, persists, and its removal results in the failure of $f_\text{xc}$ to manifest the desired excitation in $\chi$. Underneath these divergences, as demonstrated in the main text, $f_\text{xc}$ is remarkably adiabatic. Of course, these results do not preclude that a sufficient modification of the non-divergent components of $f_\text{xc}$ could manifest the desired optical peaks, but it is not clear how to construct such an $f_\text{xc}$ in principle, much less in practice.

\subsection{Quantum harmonic oscillator}

\begin{figure}[h!]
\begin{center}
\includegraphics[width=3in]{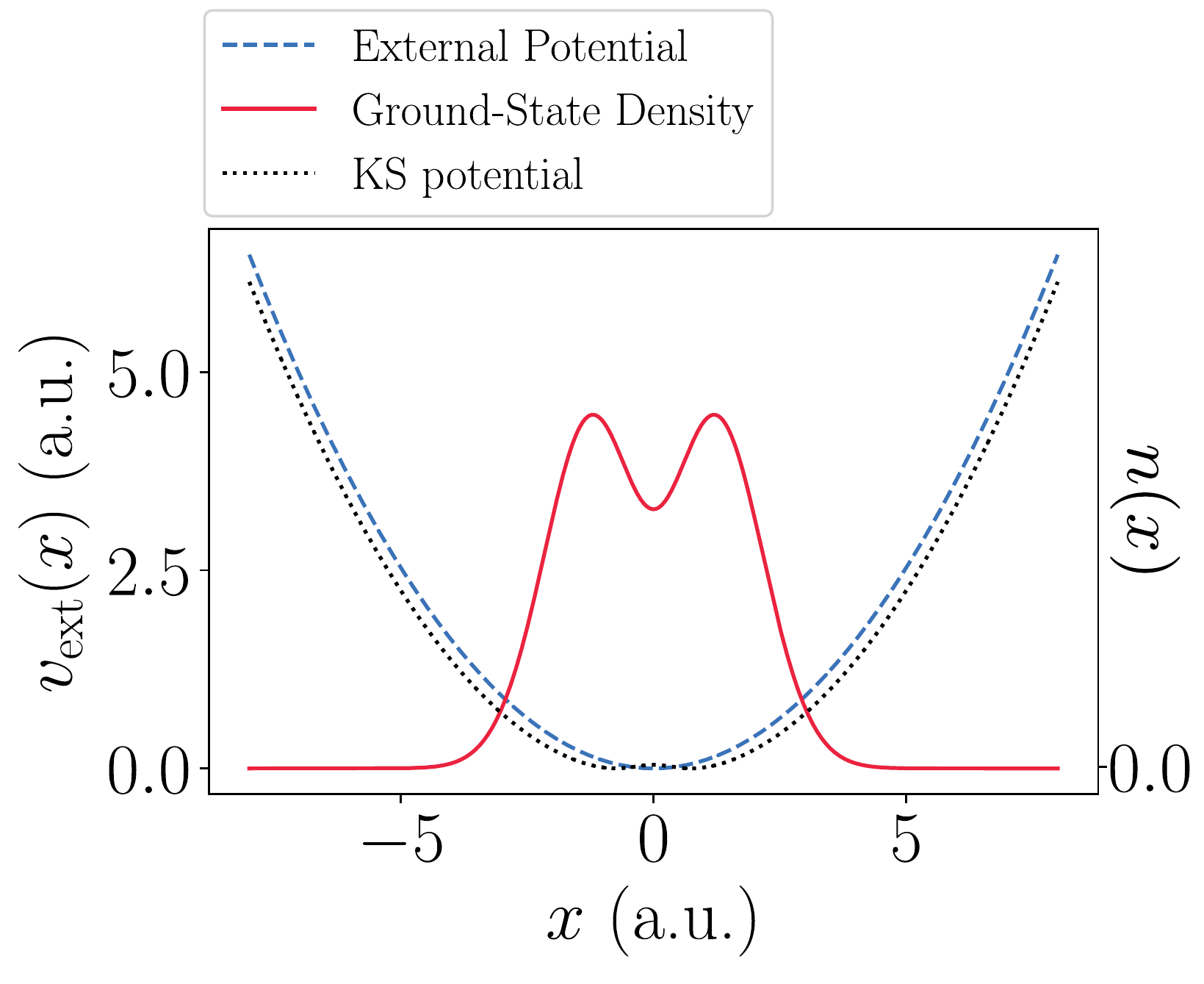}
\end{center}
\caption{The ground-state density, external potential, and reverse-engineered Kohn-Sham potential for the quantum harmonic oscillator. The external and Kohn-Sham potential have been shifted for illustrative purposes.}
\label{fig:qho}
\end{figure}

The quantum harmonic oscillator is defined with the external potential
\begin{align}
v_\text{ext}(x) = \frac{1}{2} 0.45^2 x^2.
\end{align}
The electrons are confined to the region $x \in [-8,8]$ a.u., discretized over $N_x=151$ grid points. The interacting electron density, and its corresponding Kohn-Sham potential calculated with $\mathcal{O}(10^{-15})$ error, are given in Fig$.$ \ref{fig:qho}. The response functions are calculated using the Lehman representation for $\omega \in [0,6]$ a.u., discretized over $N_\omega = 2000$ grid points. In order to construct $f_\text{xc}$ for the quantum harmonic oscillator, the eigenspace truncation parameter is $\bar{\lambda} = 10^{-9}$, and the real-space truncation parameter is $b=6.4$ a.u. The inner region in the quantum harmonic oscillator is \textit{larger} than in the case of the atom because the density is more diffuse across the domain, despite its rapid (Gaussian) decay toward the edge of the domain. 

\subsection{Gauge freedom}

The following transformation to $f_\text{xc}$ leaves the response function output from the Dyson equation, $\chi_\text{Dyson}$, unchanged,
\begin{align}
f_\text{xc}(x,x',\omega) \rightarrow f_\text{xc}(x,x',\omega) + g(x,\omega) + h(x',\omega) + c(\omega). \nonumber
\end{align}
Consider a system for which we have defined two xc kernels, $f_\text{xc}^A(\omega)$ and $f_\text{xc}^B(\omega)$ -- one of these may be the exact xc kernel. The \textit{optimal gauge} is defined as the gauge that brings $f_\text{xc}^A$ as close as possible to $f_\text{xc}^B$ in the \textit{Frobenius norm} \footnote{The Frobenius norm of a matrix $X$ defined as $||X||_F = \sum_{ij} |X_{ij}|^2$.},
\begin{align}
\min_{g,h,c} || M(g,h,c) ||_F \coloneqq \min_{g,h,c} || f_\text{xc}^A - f_\text{xc}^B - g - h - c||_F \nonumber,
\end{align}
at a given $\omega$. In a finite basis, the matrix $M$ has entries 
\begin{align}
M_{ij} = f_{\text{xc}, ij}^A - f_{\text{xc},ij}^B - h_i - g_j - c, \nonumber
\end{align}
which makes it apparent that the the gauge specifies $2N+1$ degrees of freedom out of a total $N^2$. The minimization problem above admits a closed-form solution for these degrees of freedom. In order to derive the solution to this minimization problem, first note that the optimal shift $c$ can be absorbed into the definition of $g$ and $h$. Furthermore, one can observe that, since the xc kernels are always symmetric $f_\text{xc}(x,x') = f_\text{xc}(x',x)$, it will never be favorable, from the perspective of the Frobenius norm error, to break this symmetry with the gauge. Thus, we can focus on solving the optimization problem with the matrix $M$ now having components 
\begin{align}
M_{ij} = f_{\text{xc}, ij}^A - f_{\text{xc},ij}^B - \tilde{g}_i - \tilde{g}_j, \nonumber
\end{align}
where $\tilde{g}$ is the function $g$ with the constant $c$ absorbed. The solution of this (convex) optimization problem is obtained by taking the derivative of $M$ with respect to $\tilde{g}$ and setting it equal to zero,
\begin{align}
\tilde{g}_i = \frac{1}{N} \sum_j f_{\text{xc}, ij}^A - f_{\text{xc},ij}^B - \frac{1}{2N^2} \sum_{ij} f_{\text{xc}, ij}^A - f_{\text{xc},ij}^B, \label{eq:optimalGauge}
\end{align}
where $N$ is the number of grid points. 

\begin{figure}[h!]
\begin{center}
\includegraphics[width=1.6in]{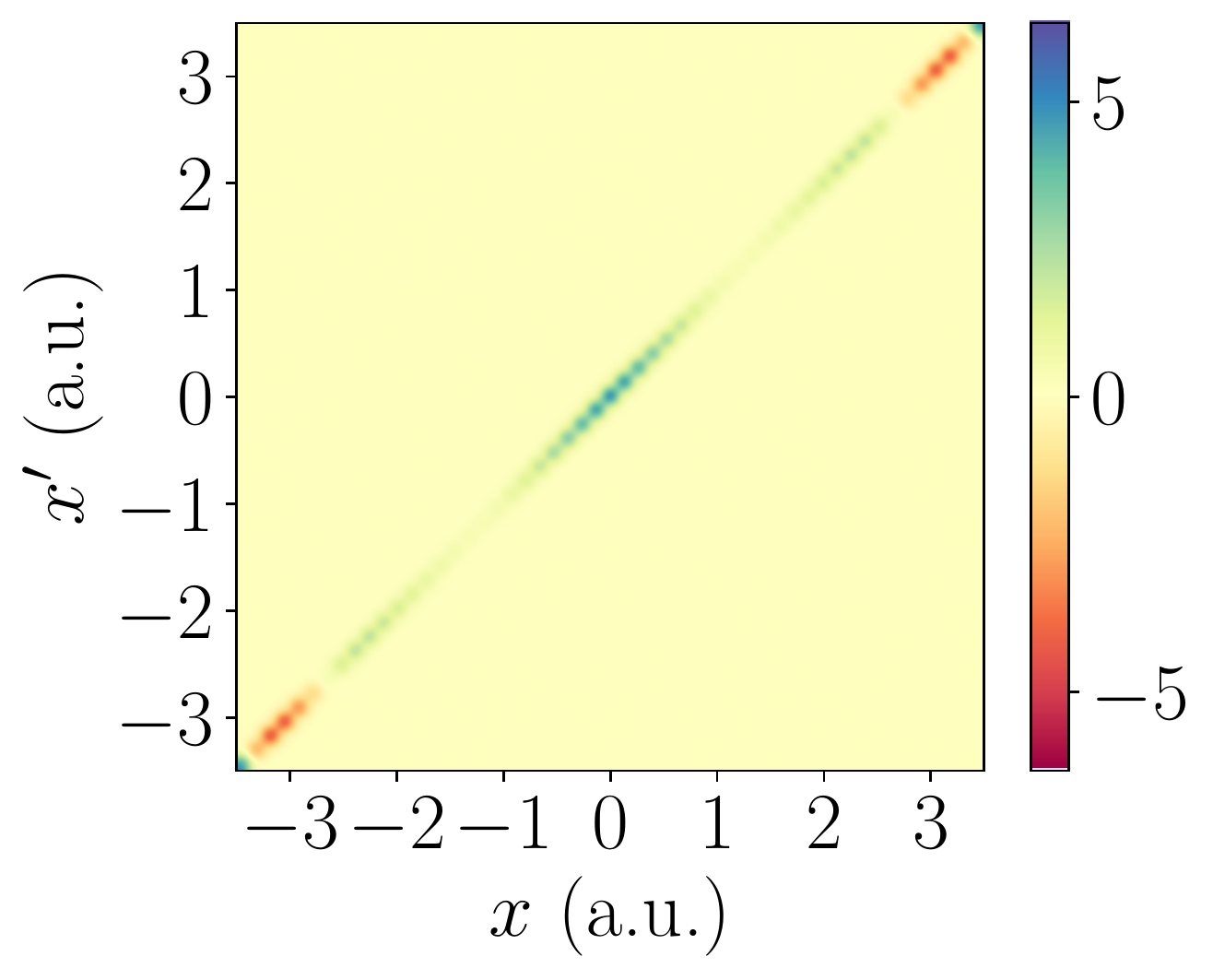}
\includegraphics[width=1.67in]{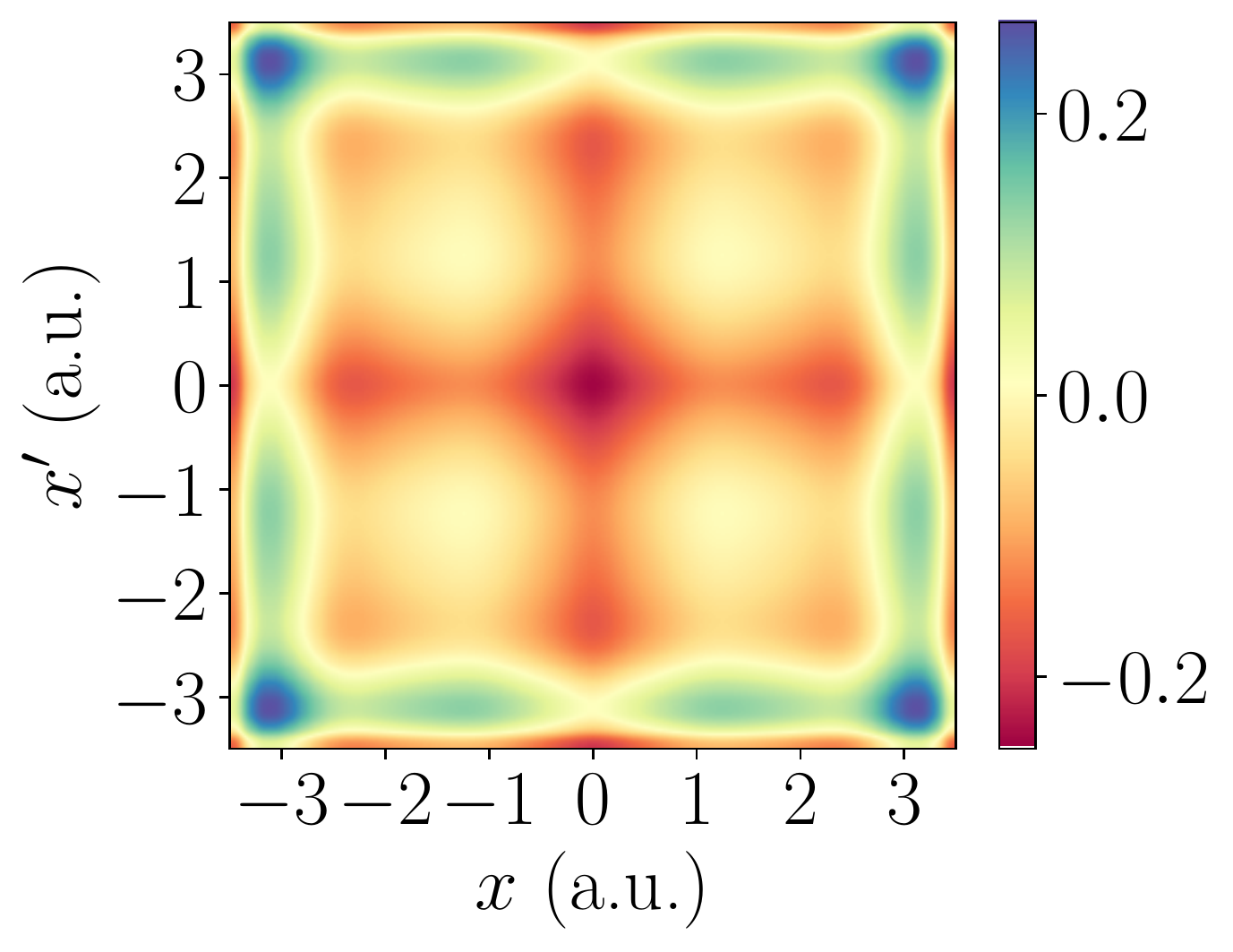}
\end{center}
\caption{The adiabatic LDA xc kernel $f_\text{xc}^\text{ALDA}$ for the atomic system (left) remains structurally dissimilar to the RPA xc kernel with the optimal gauge transform applied, $f_\text{xc}^\text{RPA} - g - h - c$ (right).}
\label{fig:gaugeRPAvsALDA}
\end{figure}

\begin{figure}[h]
\begin{center}
\includegraphics[width=3.5in]{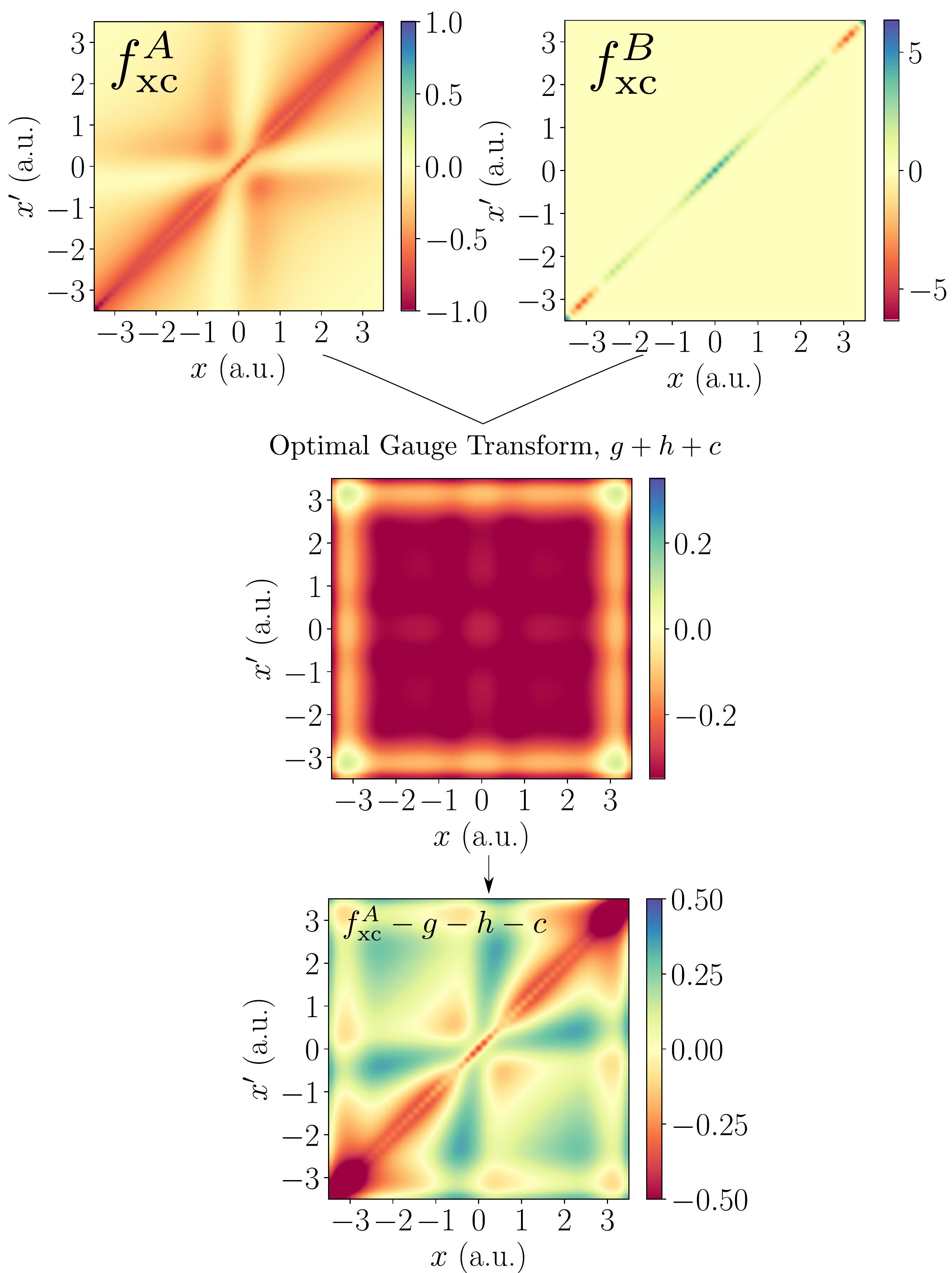}
\end{center}
\caption{An attempt is made to transform the exact adiabatic $f_\text{xc}$ (upper left) toward the adiabatic LDA xc kernel (upper right) using the optimal gauge (middle). The $f_\text{xc}$ that is closest in gauge to the adiabatic LDA (bottom) bears no resemblance in structure or magnitude to the adiabatic LDA.}
\label{fig:gaugeLDAvsAE}
\end{figure}

The examples in the main text state that $f_\text{xc}$, and its various approximations, in general are not related using a gauge transform. In the context of the atom, the first example considered is the RPA, $f_\text{xc}^A \coloneqq f_\text{xc}^\text{RPA}$, matched with the adiabatic LDA used in this work $f_\text{xc}^B \coloneqq f_\text{xc}^\text{ALDA}$ \citep{Entwistle2018}. The depiction of the optimal gauge for this example is given in Fig$.$ \ref{fig:gaugeRPAvsALDA}, in which the conclusions of the main text can be seen clearly. The optimal gauge to transform the $f_\text{xc}^\text{ALDA}$ toward the exact adiabatic $f_\text{xc}$ exhibits similar issues, Fig$.$ \ref{fig:gaugeLDAvsAE}.

\begin{figure}[H]
\begin{center}
\includegraphics[width=3in]{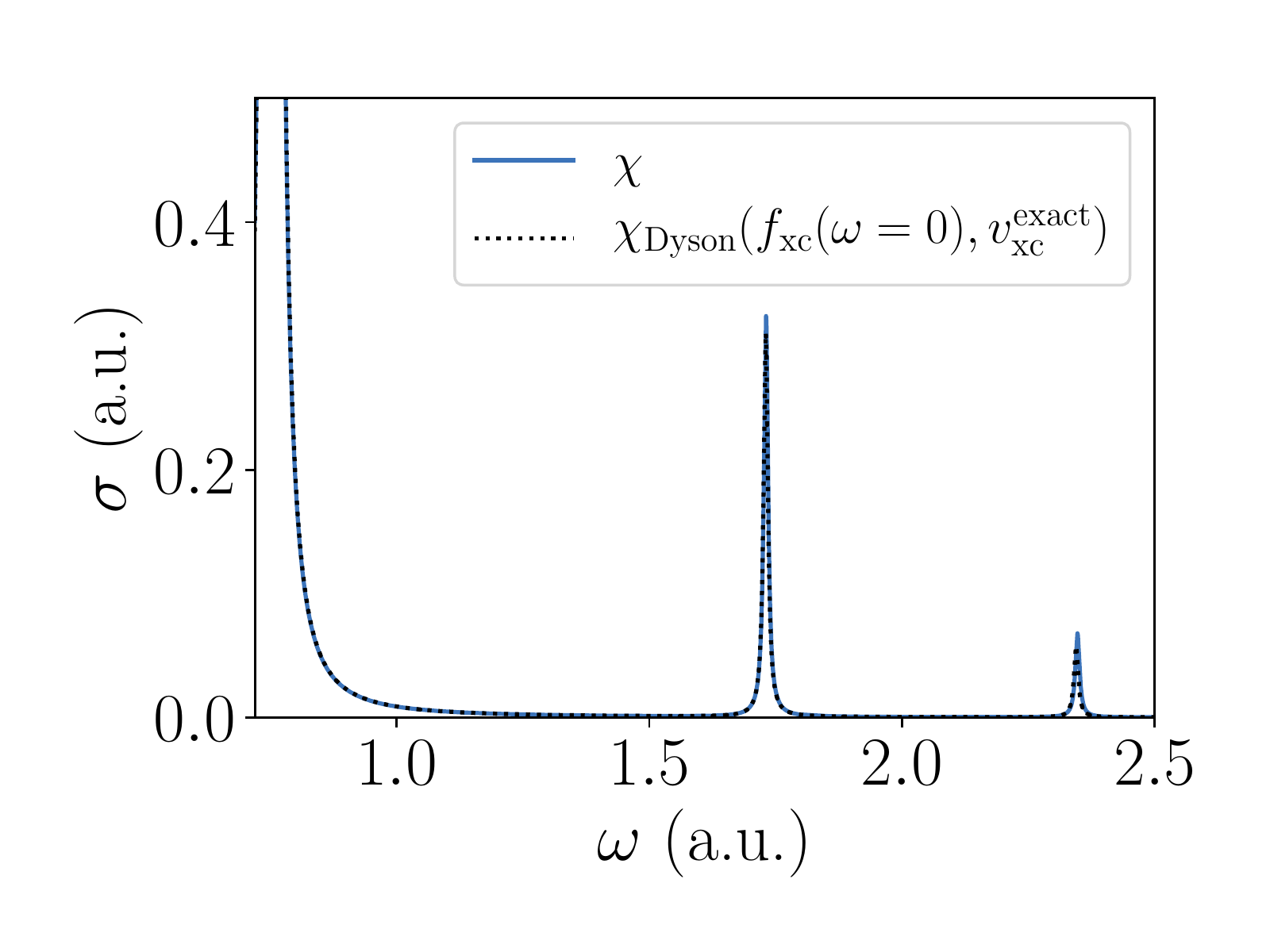}
\end{center}
\caption{The optical spectrum for the atom is computed using the interacting response function $\chi$ (blue solid), and the output of the Dyson equation $\chi_\text{Dyson}$ with the exact adiabatic xc kernel $f_\text{xc}(x,x',\omega=0)$ (black dot). The optical spectrum is reproduced accurately for the excitations shown, despite significant non-adiabatic behavior in $f_\text{xc}(\omega)$ beyond $\omega \approx 2$ a.u. The accuracy of the exact adiabatic approximation is explained with the gauge freedom inherent to $f_\text{xc}$.}
\label{fig:pseudoatomThirdExcitationOAS}
\end{figure}

\begin{figure}[H]
\begin{center}
\includegraphics[width=2.2in]{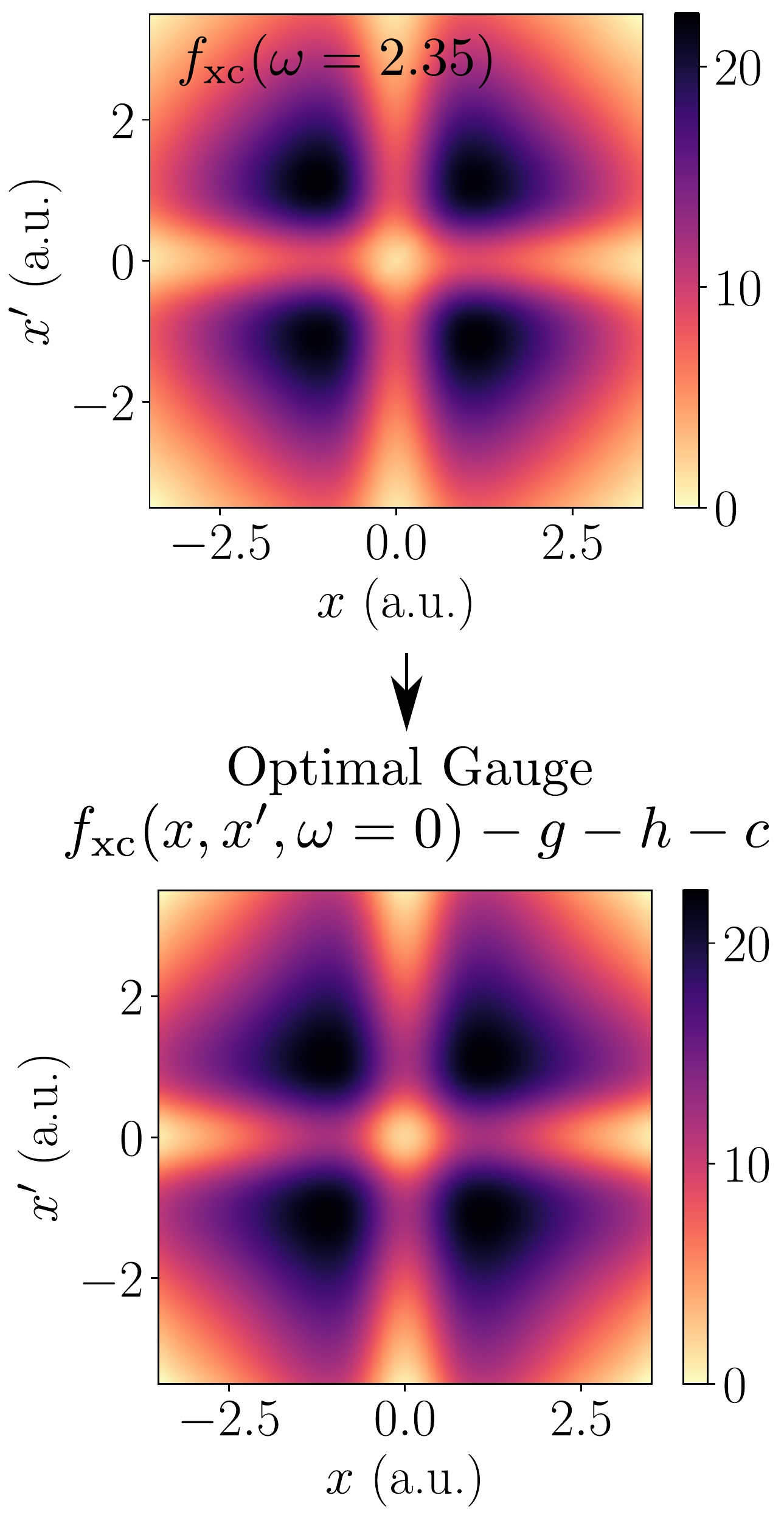}
\end{center}
\caption{The exact $f_\text{xc}$ of the atom around the third excitation, $f_\text{xc}(x,x',\omega=2.35)$ (upper panel), is the target of the optimal gauge applied to the exact adiabatic approximation, $f_\text{xc}(x,x',\omega=0)$ (lower panel). The structure of these is similar, meaning the divergence in $f_\text{xc}$ (upper) is captured excellently with the optimal gauge (lower).}
\label{fig:fxcRemoveDivergenceGauge}
\end{figure}

An interesting exception to these conclusions is given around the third excitation in the atom. Namely, the onset of a divergence occurs before the third visible excitation $\omega = 2.35$ a.u., and yet the exact adiabatic approximation captures this excitation well, see Fig$.$ \ref{fig:pseudoatomThirdExcitationOAS}. The upper panel of Fig$.$ \ref{fig:fxcRemoveDivergenceGauge} demonstrates $f_\text{xc}$ at the third excitation, which has the structure of an outer product $|u\rangle \langle u|$, as expected around a divergence. The optimal gauge, Eq$.$ (\ref{eq:optimalGauge}), to transform the exact adiabatic xc kernel, $f_\text{xc}^A = f_\text{xc}(x,x',\omega=0)$, into the exact xc kernel at the third excitation, $f_\text{xc}^B = f_\text{xc}(x,x',\omega=2.35)$, is shown in the lower panel of Fig$.$ \ref{fig:fxcRemoveDivergenceGauge}. There is a striking resemblance between the two, meaning the exact adiabatic $f_\text{xc}$ and the diverging $f_\text{xc}$ around the third excitation exist within the same family of xc kernels defined by the gauge freedom. As discussed in the main text, such an observation could account for the performance of certain adiabatic kernels beyond their expected domain of applicability.

\bibliography{suppReferences}